\documentclass[structabstract]{aa}  
\usepackage{graphicx}
\usepackage{txfonts}
\usepackage{longtable}

\begin{document}
   \title{VLTI/AMBER observations of cold giant stars: atmospheric structures and fundamental parameters \thanks{Based on observations made with the VLT Interferometer (VLTI) at Paranal Observatory under programme ID 089.D-0801}}

   \titlerunning{VLTI/AMBER observations of cool giants stars}

   \subtitle{}

   \author{B. Arroyo-Torres \inst{1} \and
		  I. Mart\'i-Vidal \inst{2} \and 
		  J. M. Marcaide \inst{1,3} \and
		  M. Wittkowski \inst{4} \and  
		  J.C. Guirado \inst{1,5} \and
		  P. H. Hauschildt \inst{6} \and
		  A. Quirrenbach \inst{7} \and
		  J. Fabregat \inst{5}
          }

   \institute{Dpt. Astronomia i Astrof\' isica, Universitat de Val\`encia, 
C/ Dr. Moliner 50, 46100, Burjassot (Val\`encia), Spain \email{belen.arroyo@uv.es} 
		 \and Onsala Space Observatory, Chalmers University of Technology, Observatorievägen 90, 43992 Onsala, Sweden
		 \and Donostia International Physics Center, Paseo de Manuel Lardizabal 4, 20018 Donostia-San Sebasti\'an, Spain
         \and ESO, Karl-Schwarzschild-St. 2, 85748, Garching bei M\" unchen, Germany
         \and Observatori Astron\`omic, Universitat de Val\`encia. C/ Catedr\'atico Jos\'e Beltr\'an 2, 46980 Paterna (Val\`encia), Spain
         \and Hamburger Sternwarte, Gojenbergsweg 112, 21029, Hamburg, Germany
         \and Landessternwarte, Zentrum f\"ur Astronomie der Universit\"at Heidelberg, K\"onigstuhl 12, 69117 Heidelberg, Germany}

   \date{Received 18 December 2012 ; Accepted 22 April 2013}

  \abstract
   {}
   {The main goal of this research is to determine the angular size and the atmospheric structures of cool giant stars ($\epsilon$~Oct, $\beta$~Peg, NU~Pav, $\psi$~Peg, and $\gamma$~Hya) and to compare them with hydrostatic stellar model atmospheres, to estimate the fundamental parameters, and to obtain a better understanding of the circumstellar environment.}
   {We conducted spectro-interferometric observations of $\epsilon$~Oct, $\beta$~Peg, NU~Pav, and $\psi$~Peg in the near-infrared K band (2.13-2.47 $\mu$m), and $\gamma$~Hya (1.9-2.47 $\mu$m) with the VLTI/AMBER instrument at medium spectral resolution ($\sim$1500). To obtain the fundamental parameters, we compared our data with hydrostatic atmosphere models (PHOENIX).}
   {We estimated the Rosseland angular diameters of $\epsilon$~Oct, $\beta$~Peg, NU~Pav, $\psi$~Peg, and $\gamma$~Hya to be 11.66$\pm$1.50\,mas, 16.87$\pm$1.00\,mas, 13.03$\pm$1.75\,mas, 6.31$\pm$0.35\,mas, and 3.78$\pm$0.65\,mas, respectively. Together with distances and bolometric fluxes (obtained from the literature), we estimated radii, effective temperatures, and luminosities of our targets. In the $\beta$~Peg visibility, we observed a molecular layer of CO with a size similar to that modeled with PHOENIX. However, there is an additional slope in absorption starting around 2.3\,$\mu$m. This slope is possibly due to a shell of H$_{2}$O that is not modeled with PHOENIX (the size of the layer increases to about 5\% with respect to the near-continuum level). The visibility of $\psi$~Peg shows a low increase in the CO bands, compatible with the modeling of the PHOENIX model. The visibility data of $\epsilon$~Oct, NU~Pav, and $\gamma$~Hya show no increase in molecular bands.}
   {The spectra and visibilities predicted by the PHOENIX atmospheres agree with the spectra and the visibilities observed in our stars (except for $\beta$~Peg). This indicates that the opacity of the molecular bands is adequately included in the model, and the atmospheres of our targets have an extension similar to the modeled atmospheres. The atmosphere of $\beta$~Peg is more extended than that predicted by the model. The role of pulsations, if relevant in other cases and unmodeled by PHOENIX, therefore seems negligible for the atmospheric structures of our sample. The targets are located close to the red limits of the evolutionary tracks of the STAREVOL model, corresponding to masses between 1~$M_{\odot}$ and 3~$M_{\odot}$. The STAREVOL model fits the position of our stars in the Hertzsprung-Russell (HR)  diagram better than the Ekstr\"om model does. STAREVOL includes thermohaline mixing, unlike the Ekstr\"om model, and complements the latter for intermediate-mass stars.}


\maketitle

%

\section{Introduction}

\begin{table*}
\caption{VLTI/AMBER observations}
\centering
\begin{tabular}{lcccccccc}
\hline
\hline
Target (Sp. type)  & Date  & Baseline & Projected Baseline & PA  & Calibrator   \\
   &   & & m  & deg &  \\
\hline
$\epsilon$ Oct (M5 III)  & 2012 Jun 25  & D0-A1-C1 & 28.9/15.4/15.1 & -119/79/-138 & HIP 104755\\
 & 2012 Aug 02  & B2-A1-C1 & 10.26/15.62/7.83 & -96.1/57.8/22.7 & HIP 104755\\
 $\beta$ Peg (M2.5 II-III)  & 2012 Jun 25  & D0-A1-C1 & 30.4/15.6/16.4 & -122/76.6/140 & HIP 114144 - HIP 1168 \\
 & 2012 Aug 09  & B2-A1-C1 & 11.1/14.5/7.0 & -69.2/82.80/35.20 & HIP 114144 - HIP 1168\\
 NU Pav  (M6 III) & 2012 Aug 02  & B2-A1-C1 & 10.9/15.8/9.8 & -70.6/71.6/28.8 & HIP 82363 \\
 $\psi$ Peg  (M3 III) & 2012 Jun 16  & D0-I1-G1 & 82.1/32.8/66.1 & 102/-128/124.5 & HIP 114144 - HIP 1168\\
$\gamma$~Hya  (G8 III) & 2013 Mar 16 & A1-G1-J3 & 74.9/132.2/135.7 & 118/15.3/47.7 & K Hya\\
\hline
\end{tabular}
\tablefoot{Details of our observations. The AMBER instrument mode is K-2.3\,$\mu$m (2.12-2.47\,$\mu$m). The projected baseline is the projected baseline length for the AT VLTI baseline used, and PA is the position angle of the baseline (North through East). $\gamma$~Hya has been observed in K-2.1 and K-2.3 bands, which together cover the range 1.9-2.47\,$\mu$m.}
\label{Log_obs}
\end{table*}

The motivation for our study is to improve our understanding of the circumstellar environment of asymptotic giant branch stars (AGBs) close to the photosphere, to obtain estimates about their fundamental parameters, and to locate them in the Hertzsprung-Russell (HR)  diagram. The location of the stars in the HR diagram is very important for calibrating stellar evolutionary models for intermediate-mass stars.
 
Interferometric techniques at visible and IR wavelengths are important for resolving the stellar disk to better understand the circumstellar environment (Quirrenbach et al. \cite{Quirrenbach1993}, Perrin et al. \cite{Perrin2004}). Recent studies with VLTI/AMBER and VLTI/MIDI have provided information about the pulsation and the mass-loss of AGB stars (Ohnaka et al. \cite{Ohnaka2006}, \cite{Ohnaka2007}; Wittkowski et al. \cite{Wittkowski2007}; Chiavassa et al. \cite{Chiavassa2010}; Karovicova et al. \cite{Karovicova2011}, \cite{Karovicova2013}) and about the structure of the molecular distribution in AGB stars (Wittkowski et al. \cite{Wittkowski2008}, \cite{Wittkowski2011}; Mart{\'{\i}}-Vidal et al. \cite{Marti2011}).   

Quirrenbach et al. (\cite{Quirrenbach1993}, \cite{Quirrenbach2001}) studied the TiO band (around 712\,nm) in the atmosphere of cold giant stars (spectral type M). Their interferometric observations were made with two filters, one centered on the TiO band, and the other on the continuum close to that band. They observed an increase of the size of the star corresponding to the TiO band with respect to the size in the continuum. After fitting the PHOENIX models (Hauschildt \& Baron \cite{Hausch1999}) to their data, they concluded that the diameter ratio between the TiO band and the continuum agreed with models computed for a mass of 0.5\,$M_{\odot}$; but the evolutionary models predict for these stars masses of about 5\,$M_{\odot}$. This disagreement might be explained by the existence of a transition zone at the base of the stellar wind (Tsuji \cite{Tsuji2008}), which could provide sufficient opacity in the TiO band to make the AGB larger than the size predicted by the PHOENIX model. 

Mart{\'{\i}}-Vidal et al. (\cite{Marti2011}) observed RS~Cap (AGB star of spectral type M6/M7III) with the VLTI/AMBER instrument in the K band. They found that the apparent size of the star increased around 12\% in the CO band (2.29\,$\mu$m-2.47\,$\mu$m). The fit to the data with MARCS models (Gustafsson et al. \cite{Gustafsson2008})  was reasonable, although the lower visibilities in the CO band were not reproduced. These authors added an ad hoc spherical water envelope around the star (Perrin et al. \cite{Perrin2004}) that made the synthetic visibilities and the observations in the CO bands appear consistent.

Cruzal{\`e}bes et al. (\cite{Cruzalebes2013}) observed sixteen red giants and supergiants with VLTI/AMBER over a two-year period. They used MARCS models to fit their data. Their estimates of the angular diameters were moderately dependent on the variation of the model input parameters T$_{eff}$, log(g), and $\xi_{turb}$. Eight of these sources were studied for the first time but the others had been studied earlier with Long-Baseline Interferometry (LBI), and the angular diameter estimates obtained with both methods were similar.

Cusano et al. (\cite{Cusano2012}) studied five giant stars while investigating planet formation around stars more massive than the Sun. They estimated the uniform disk (UD) and limb-darkened (LD) angular diameters and the effective temperatures of these sources. The measurements of both angular diameters (UD and LD) were consistent within 1.5$\sigma$, the differences being smaller than 0.8\%. Their estimates were also consistent with the values derived by da Silva et al. \cite{Silva2006}.

In this paper we study a sample of cool giant stars with VLTI/AMBER. We locate our targets in the HR diagram and compare our results with those of the red supergiant stars studied in Arroyo-Torres et al. (\cite{Arroyo2013}). The remainder of this paper is structured as follows: in Sect. 2, we describe our AMBER observations and the data reduction, in Sect. 3, we present the PHOENIX model used, in Sect. 4, we report and discuss our results. Finally, we present our conclusions in Sect. 5.   

\section{Observations and data reduction}

We observed $\epsilon$~Oct (M5~III), $\beta$~Peg (M2.5~II-III), NU~Pav (M6~III), $\psi$~Peg (M3~III), and $\gamma$~Hya  (G8 III) with the ESO Very Large Telescope Interferometer (VLTI), using three of the auxiliary telescopes of 1.8\,m diameter, and the AMBER instrument (Astronomical Multi-BEam combineR) with the external fringe tracker FINITO (Petrov et al. \cite{Petrov2007}). We observed in medium-resolution mode (R~$\sim$~1500) in the K-2.3\,$\mu$m band ($\gamma$~Hya has been also observed in band K-2.1). We scheduled our observations as sequences of cal-sci-cal (cal is calibrator and sci is our target), with five scans for each of them. The integration time (DIT) of each frame was 0.2s (for $\epsilon$~Oct, $\beta$~Peg, NU~Pav, and $\psi$~Peg) and 1.0s (for $\gamma$~Hya). The stars $\epsilon$~Oct and $\beta$~Peg were each observed with two different arrays. In Table \ref{Log_obs}, we show information about our observations and the calibrator used for each target. In Table \ref{calibrator}, we show the calibrators used for our observations, selected from the ESO Calibration Selector CalVin, in turn based on the catalog of Lafrasse et al. (\cite{Lafrasse2010}).

\begin{figure*}
\centering
\includegraphics[width=0.49\hsize]{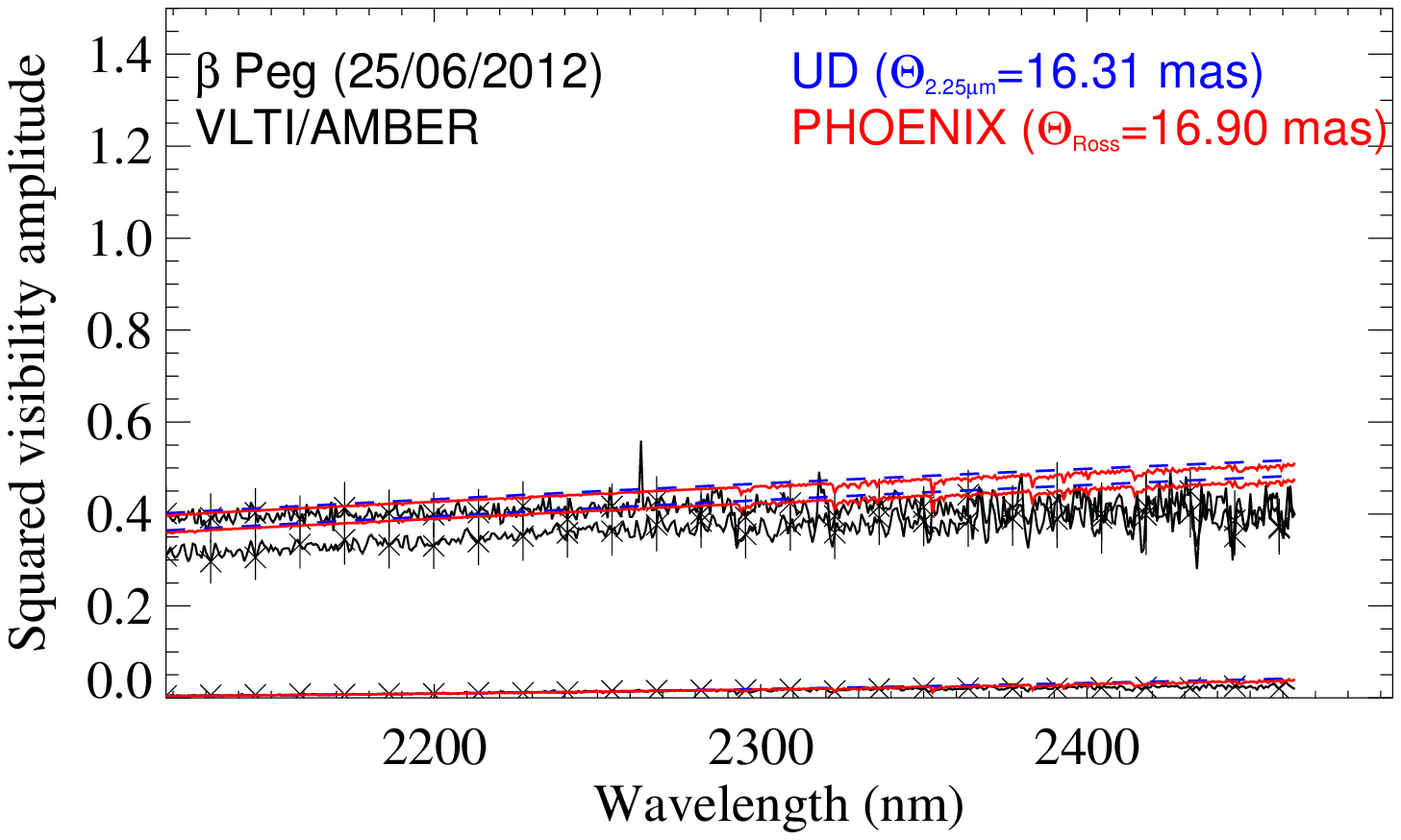}
\includegraphics[width=0.49\hsize]{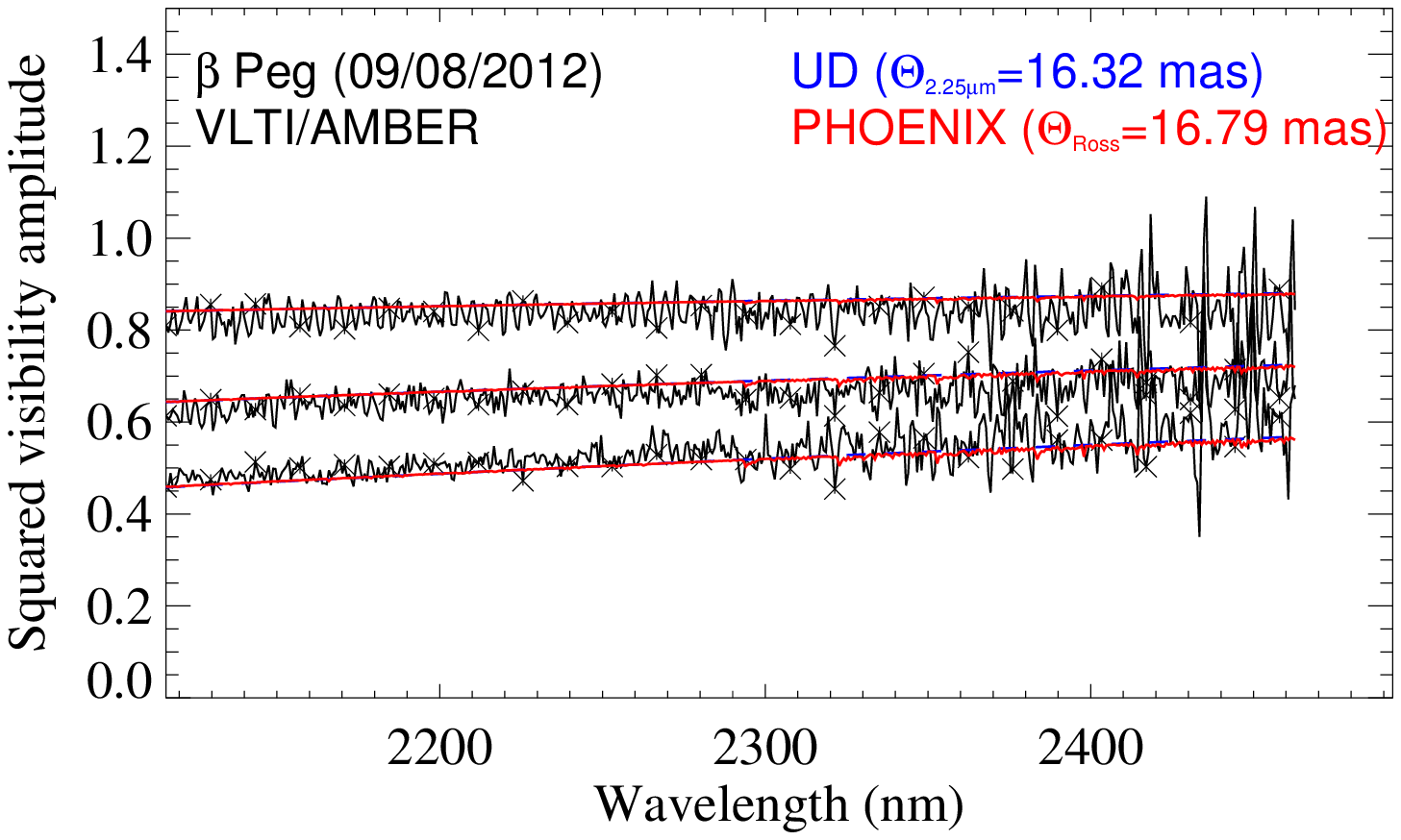}
\includegraphics[width=0.49\hsize]{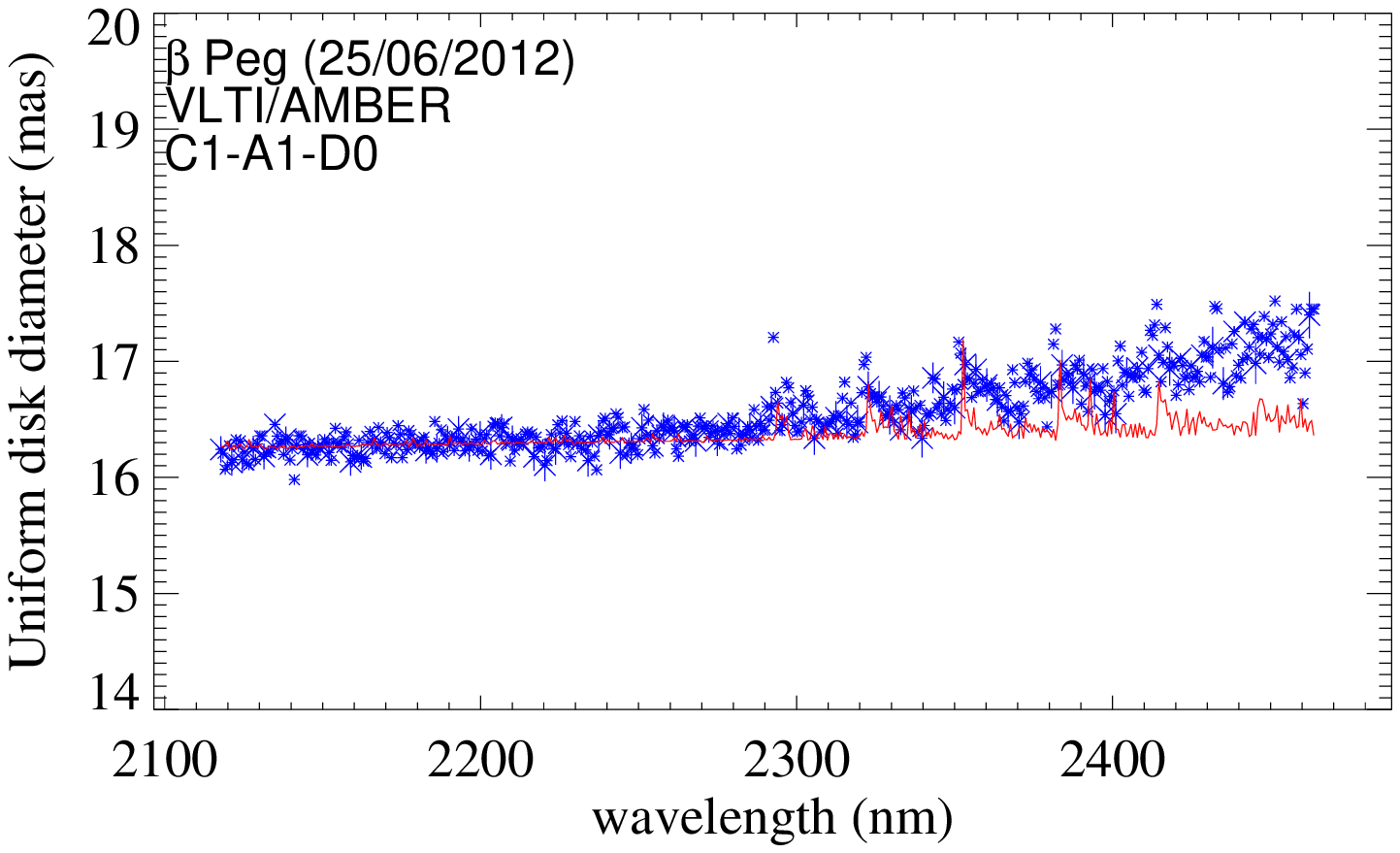}
\includegraphics[width=0.49\hsize]{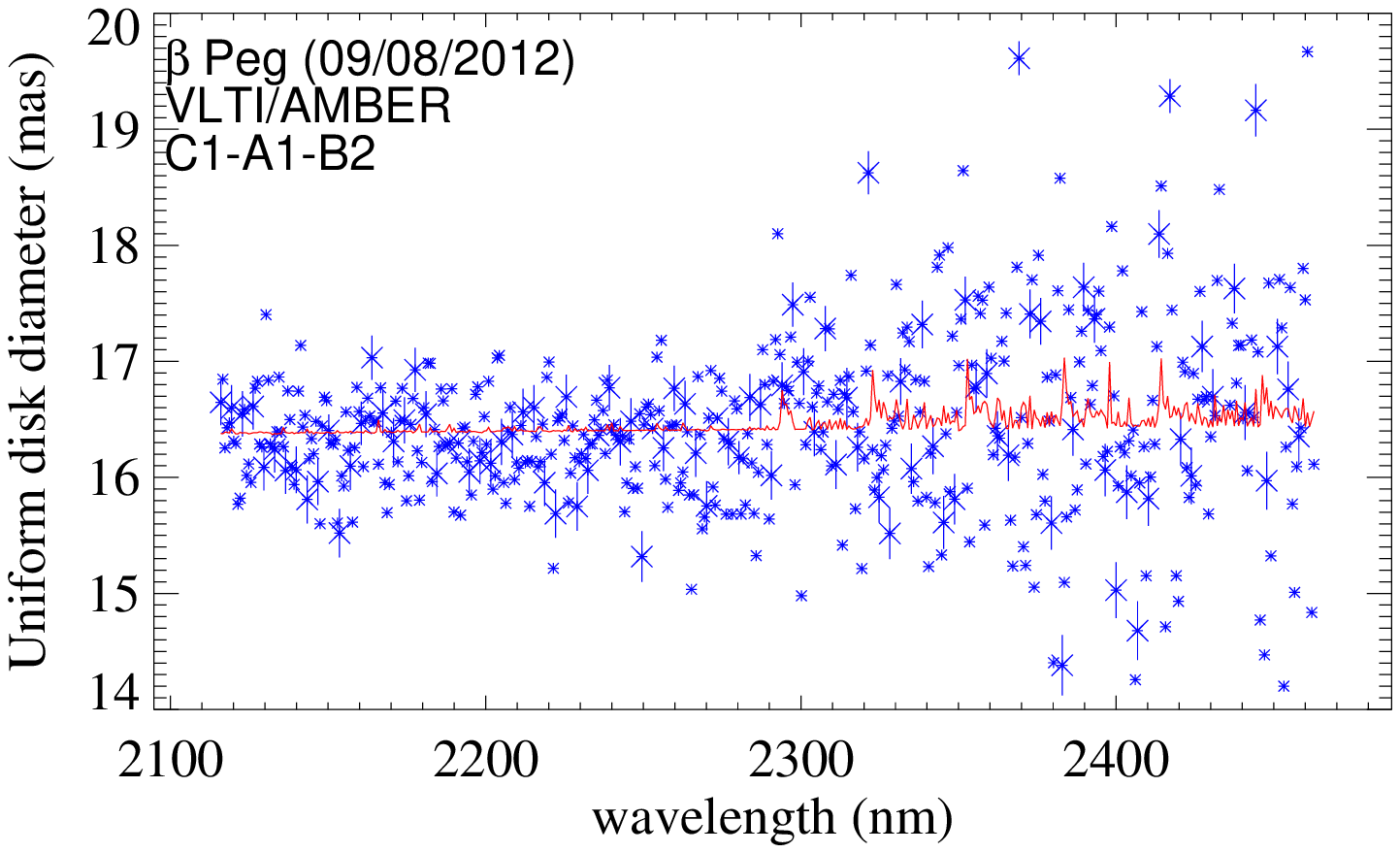}
\includegraphics[width=0.49\hsize]{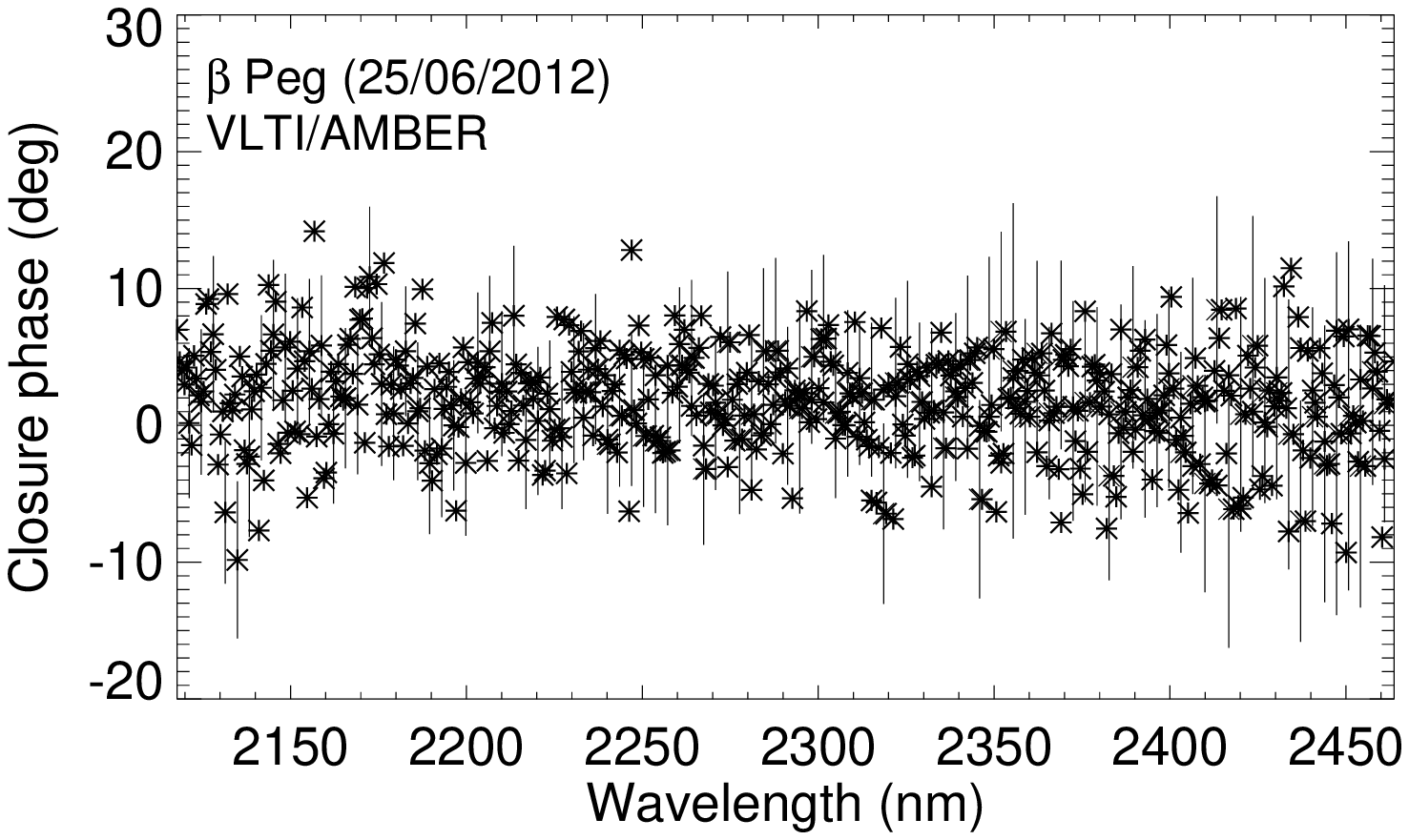}
\includegraphics[width=0.49\hsize]{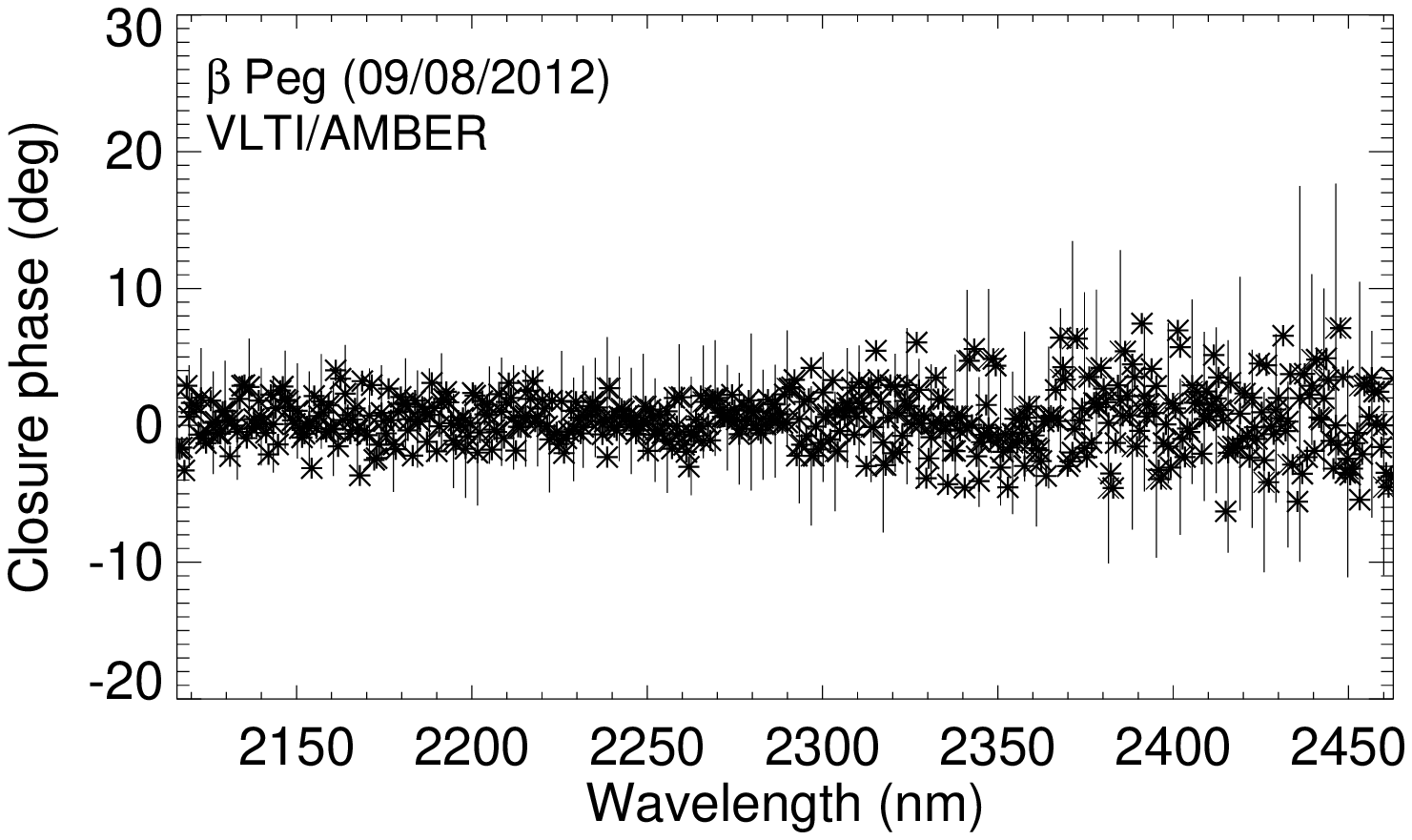}
\includegraphics[width=0.49\hsize]{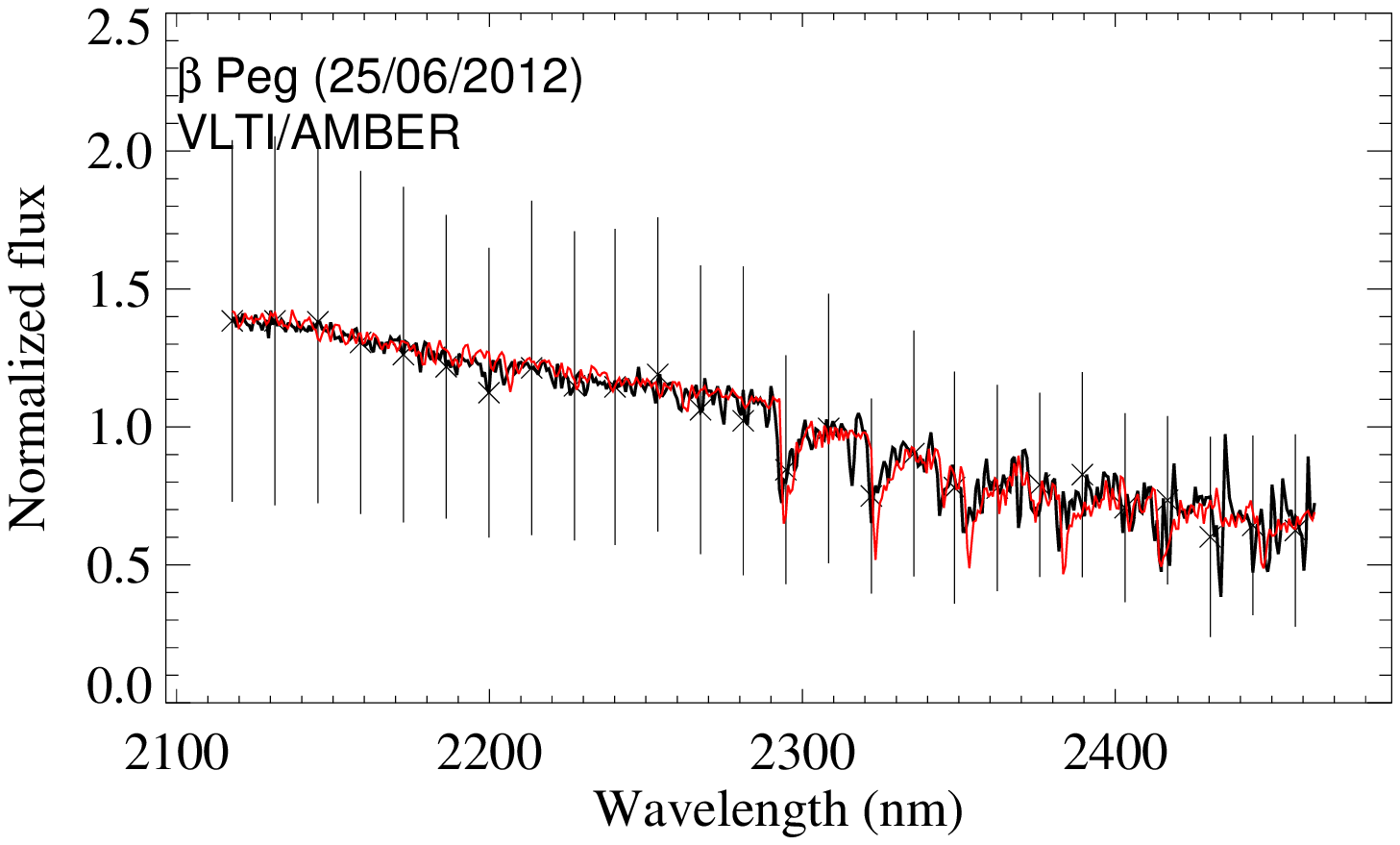}
\includegraphics[width=0.49\hsize]{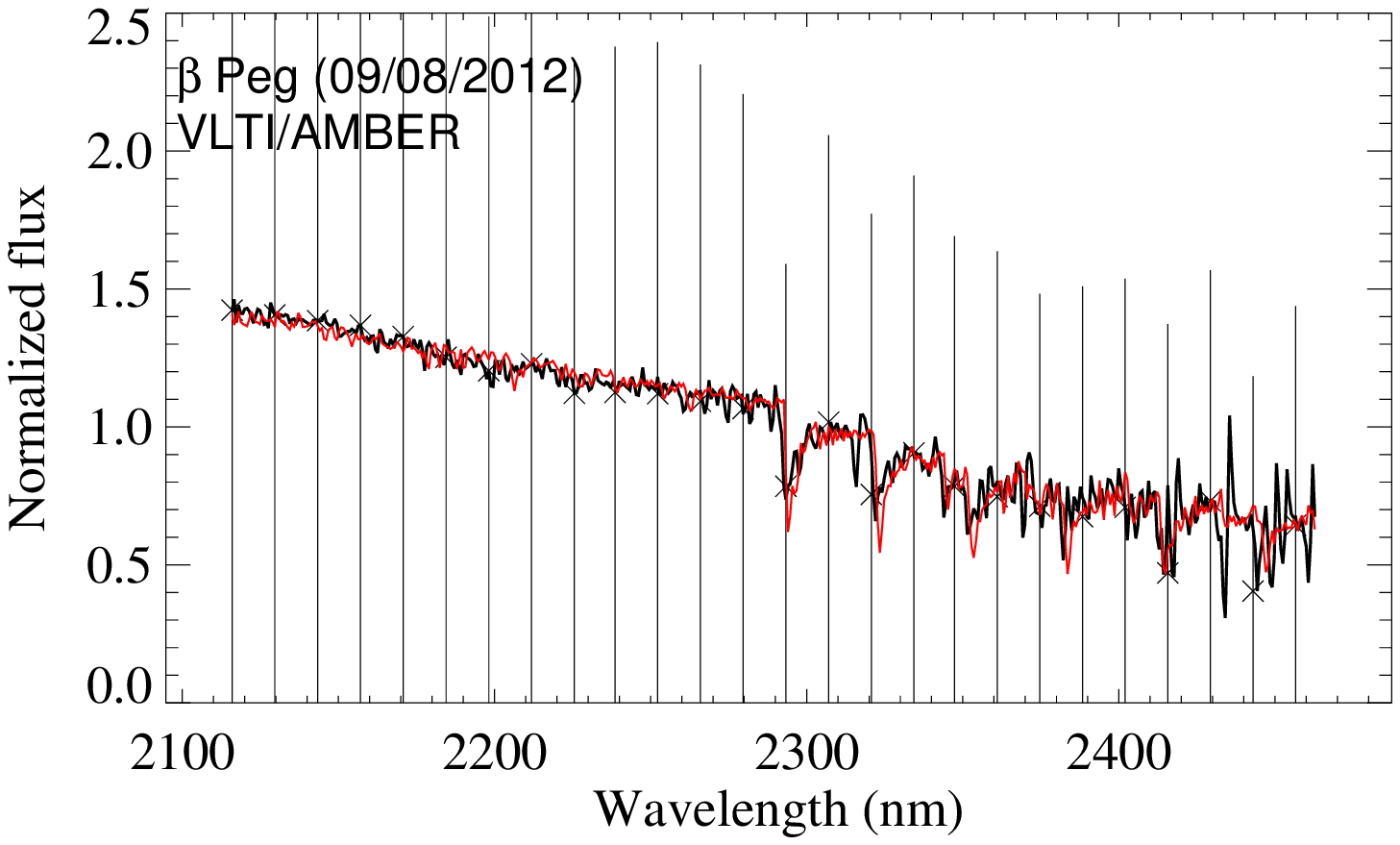}
\caption{
Left: from top to bottom, observed (black) squared visibility amplitudes, UD diameters predicted from our data (blue) and from the best-fit PHOENIX model (red), closure phases in degrees, and normalized flux of $\beta$~Peg obtained on 2012 Jun 25. Right: same as left for data obtained on 2012 Aug 09.}
\label{resul_betaPeg_fit}
\end{figure*}

\onlfig{2}{
\begin{figure*}
\centering
\includegraphics[width=0.40\hsize]{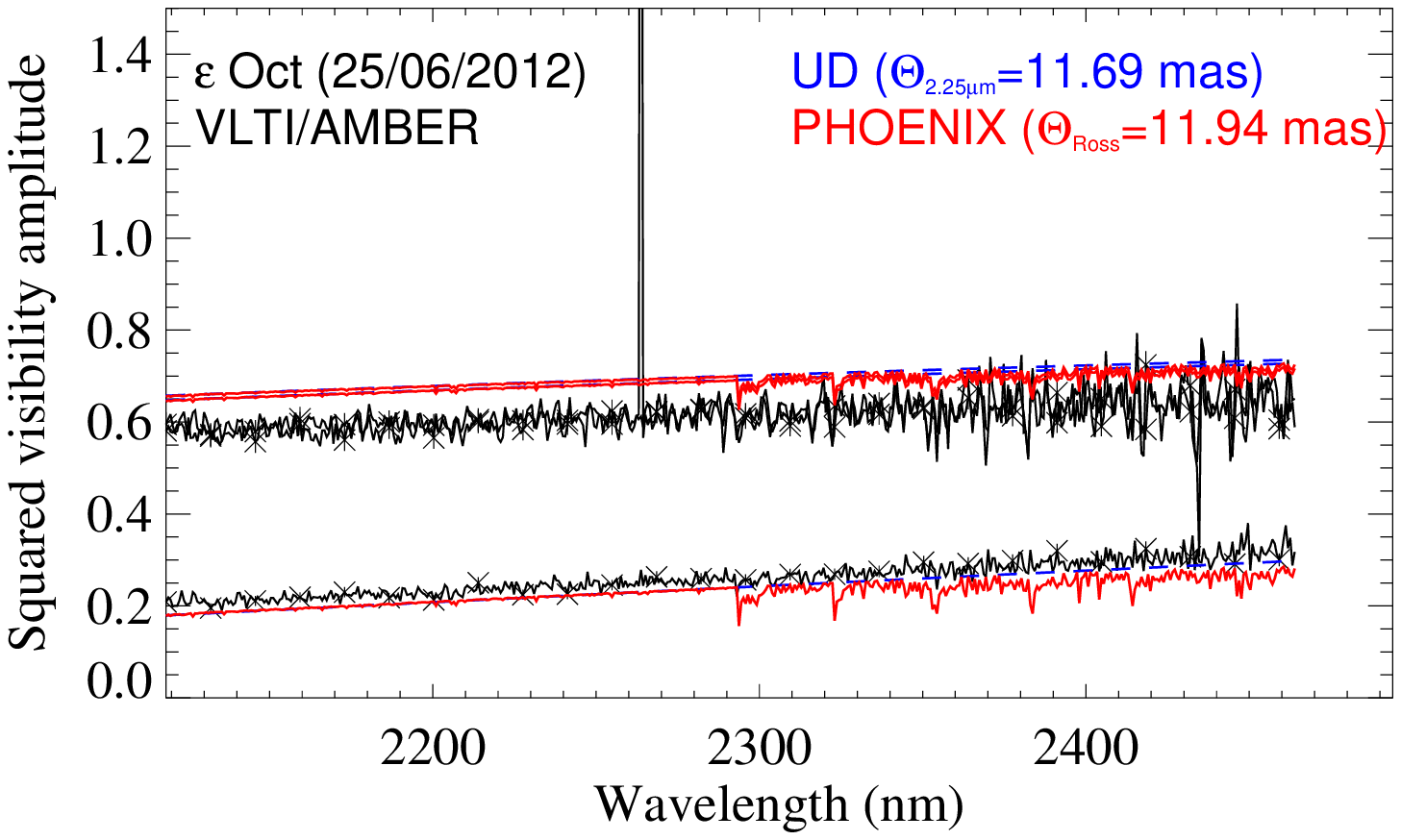}
\includegraphics[width=0.40\hsize]{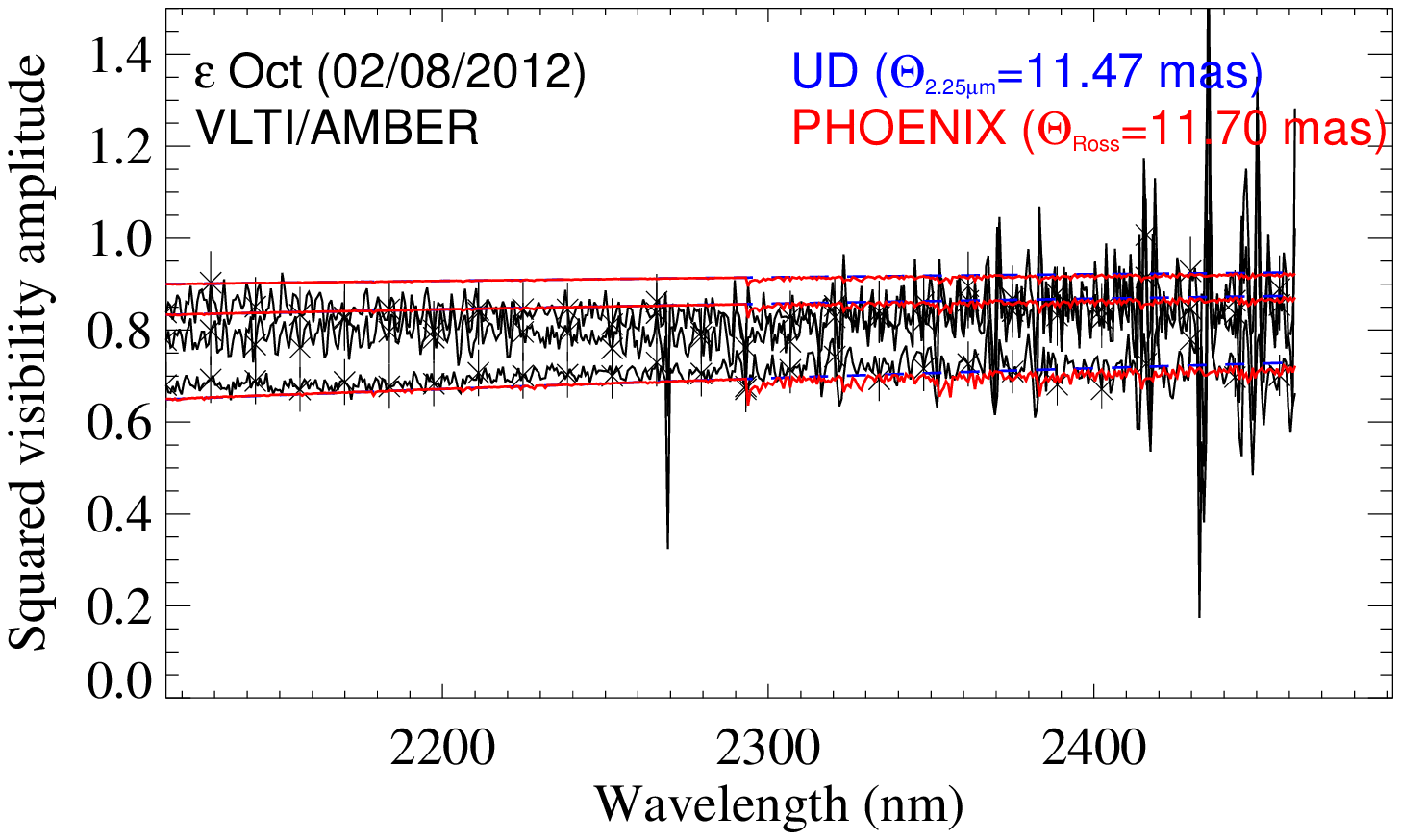}
\includegraphics[width=0.40\hsize]{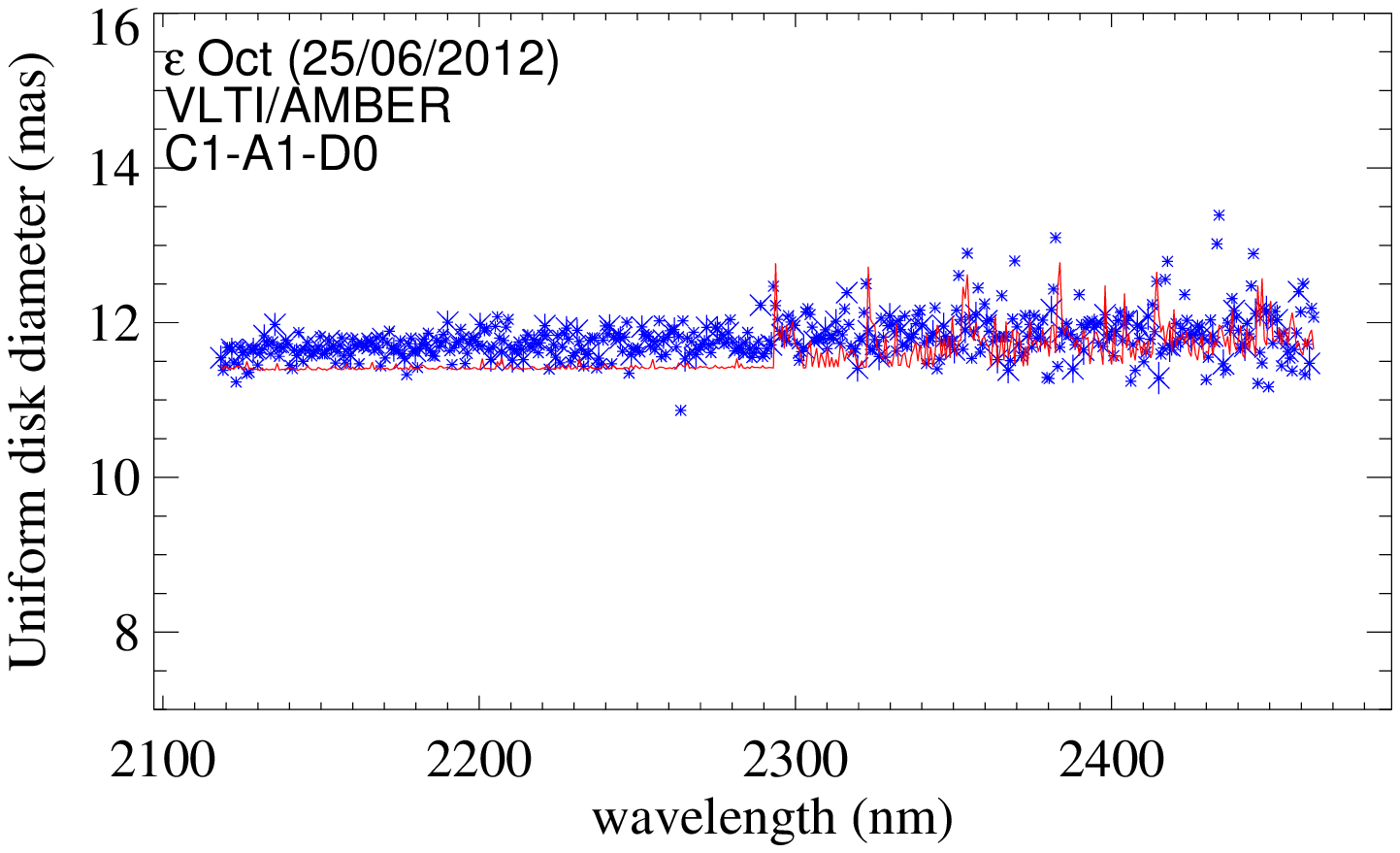}
\includegraphics[width=0.40\hsize]{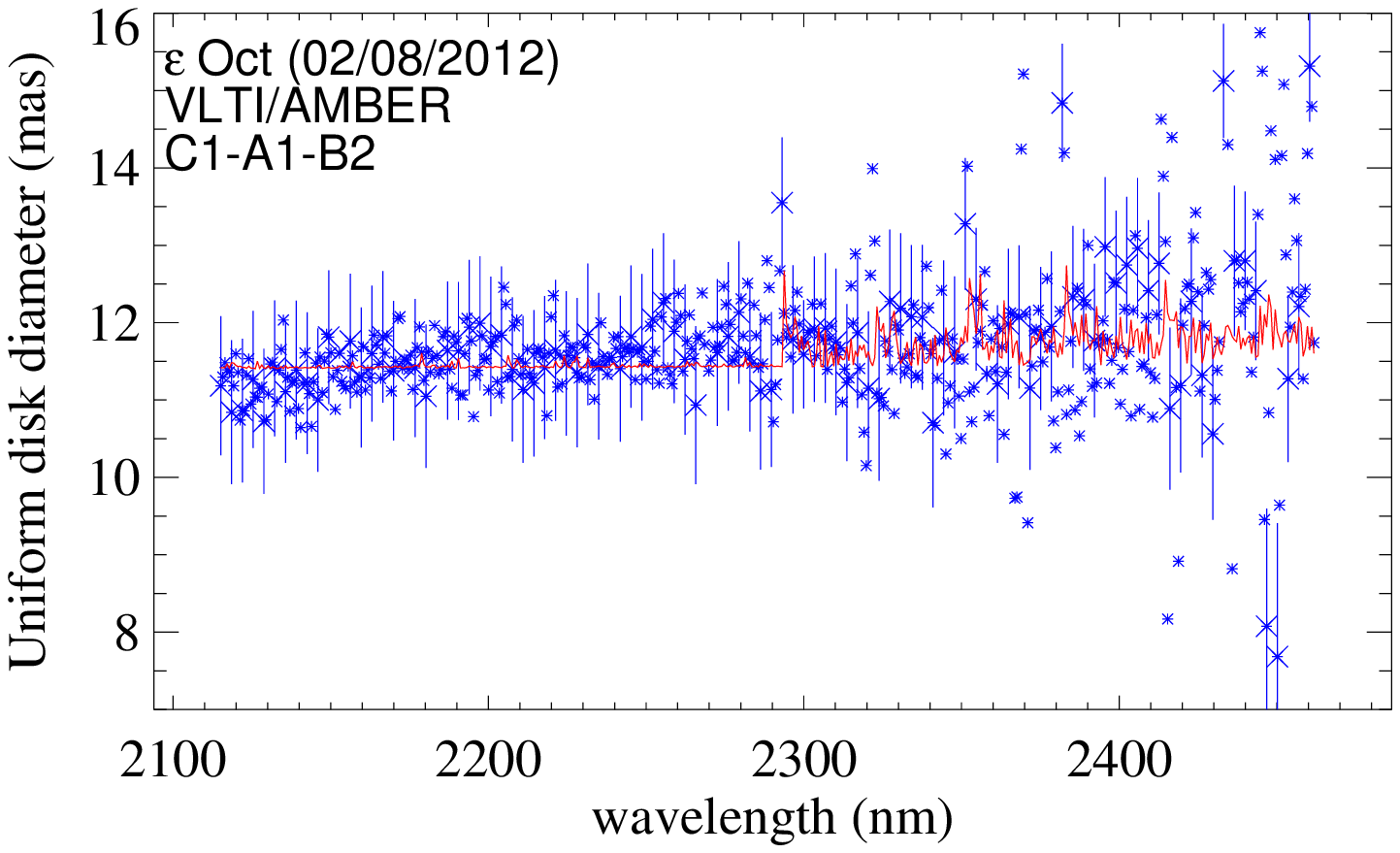}
\includegraphics[width=0.40\hsize]{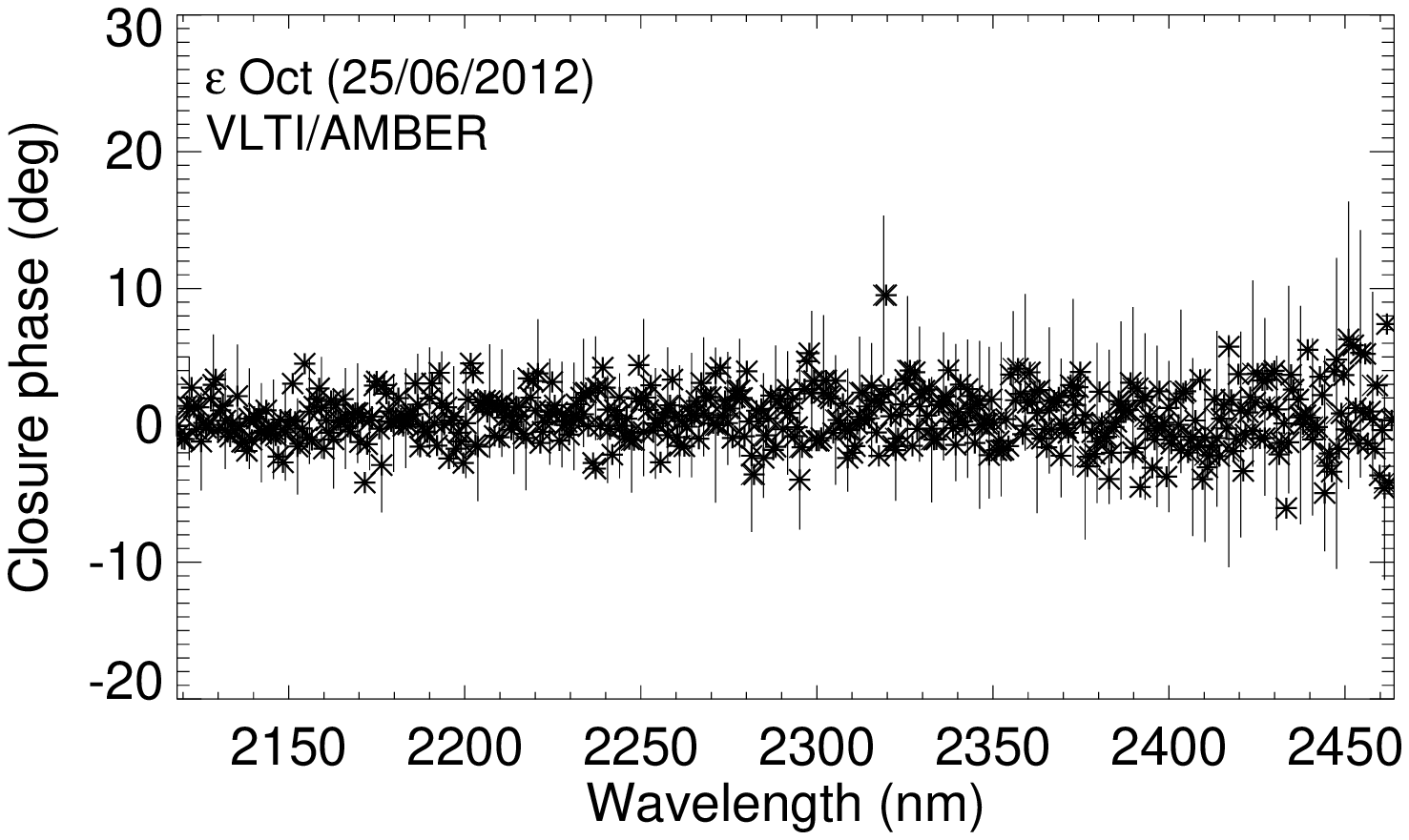}
\includegraphics[width=0.40\hsize]{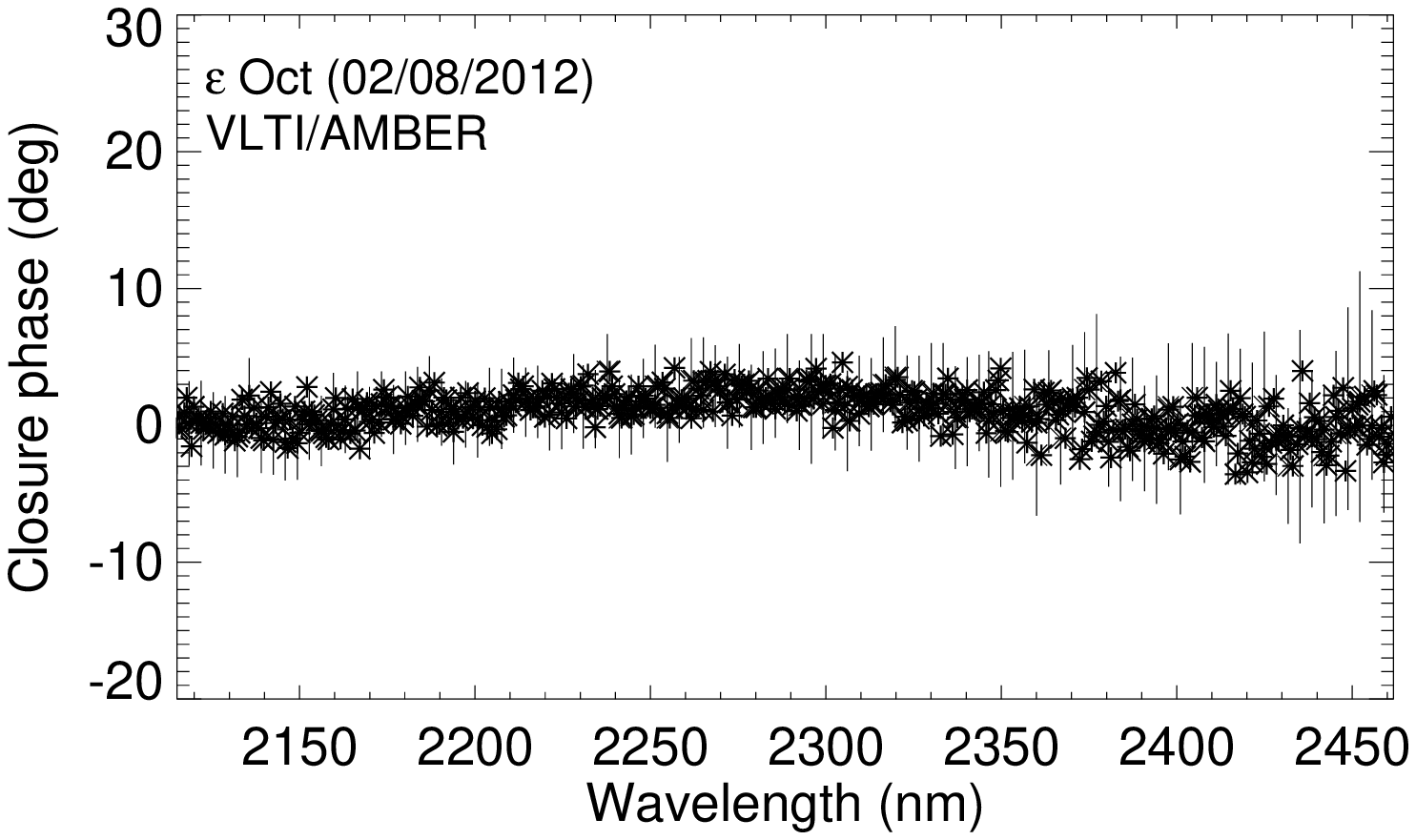}
\includegraphics[width=0.40\hsize]{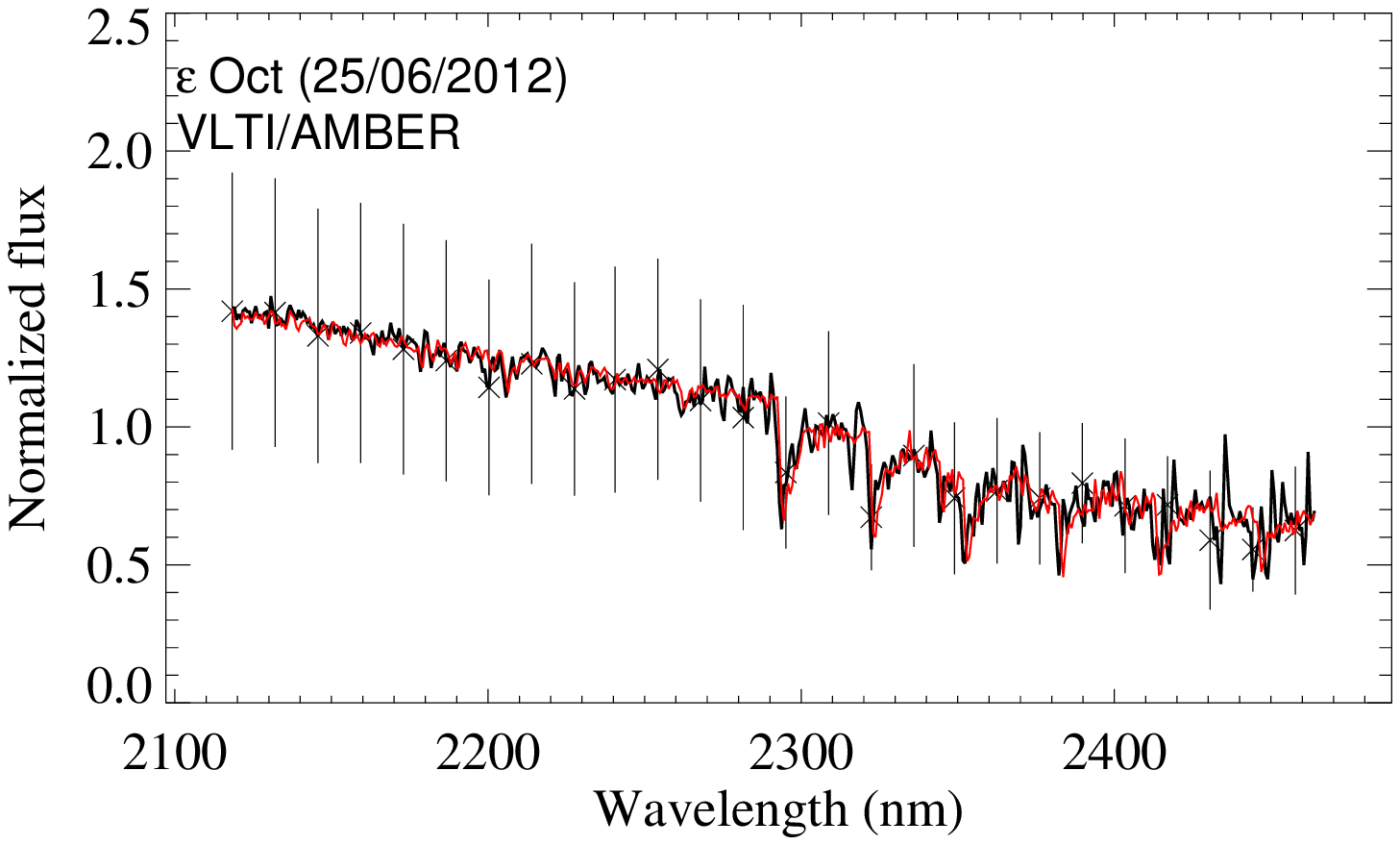}
\includegraphics[width=0.40\hsize]{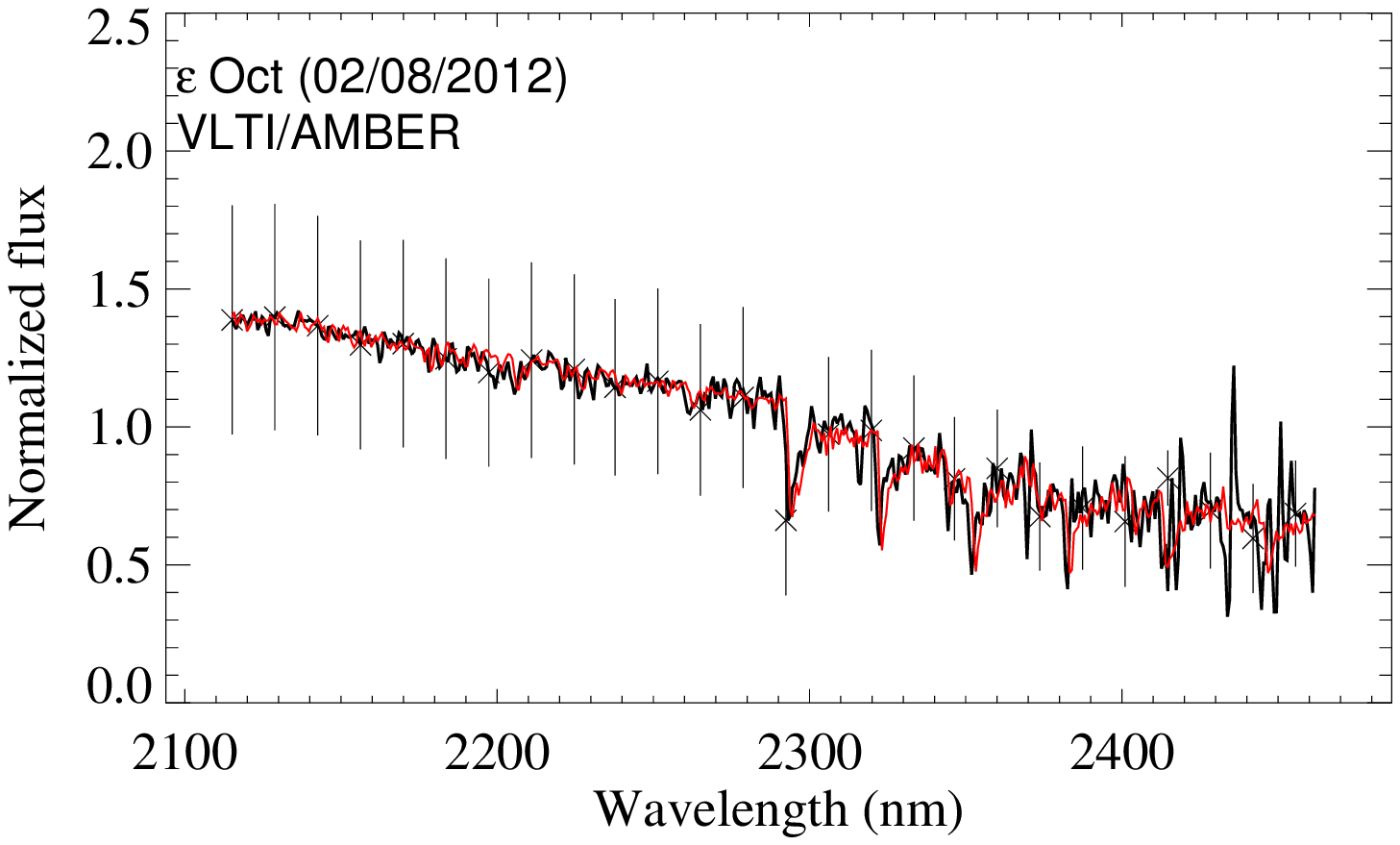}
\caption{
Left: from top to bottom, observed (black) squared visibility amplitudes, UD diameters predicted from our data (blue) and from the best-fit PHOENIX model (red), closure phases in degrees, and  normalized flux of $\epsilon$~Oct obtained on 2012 Jun 25. Right: same as left, but for data obtained on 2012 Aug 02.}
\label{resul_BoOct_fit}
\end{figure*} }

\onlfig{3}{
\begin{figure*}
\centering
\includegraphics[width=0.40\hsize]{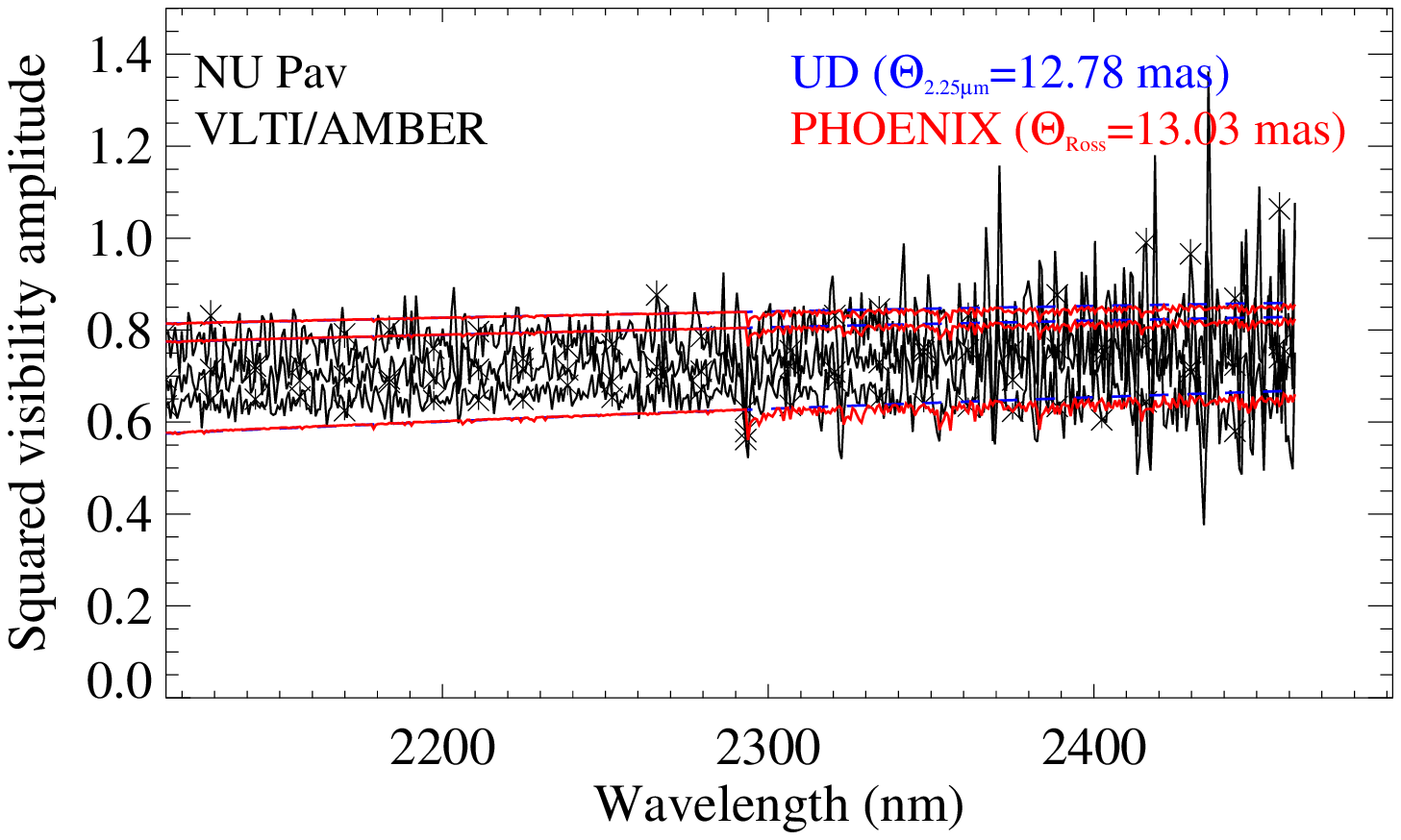}
\includegraphics[width=0.40\hsize]{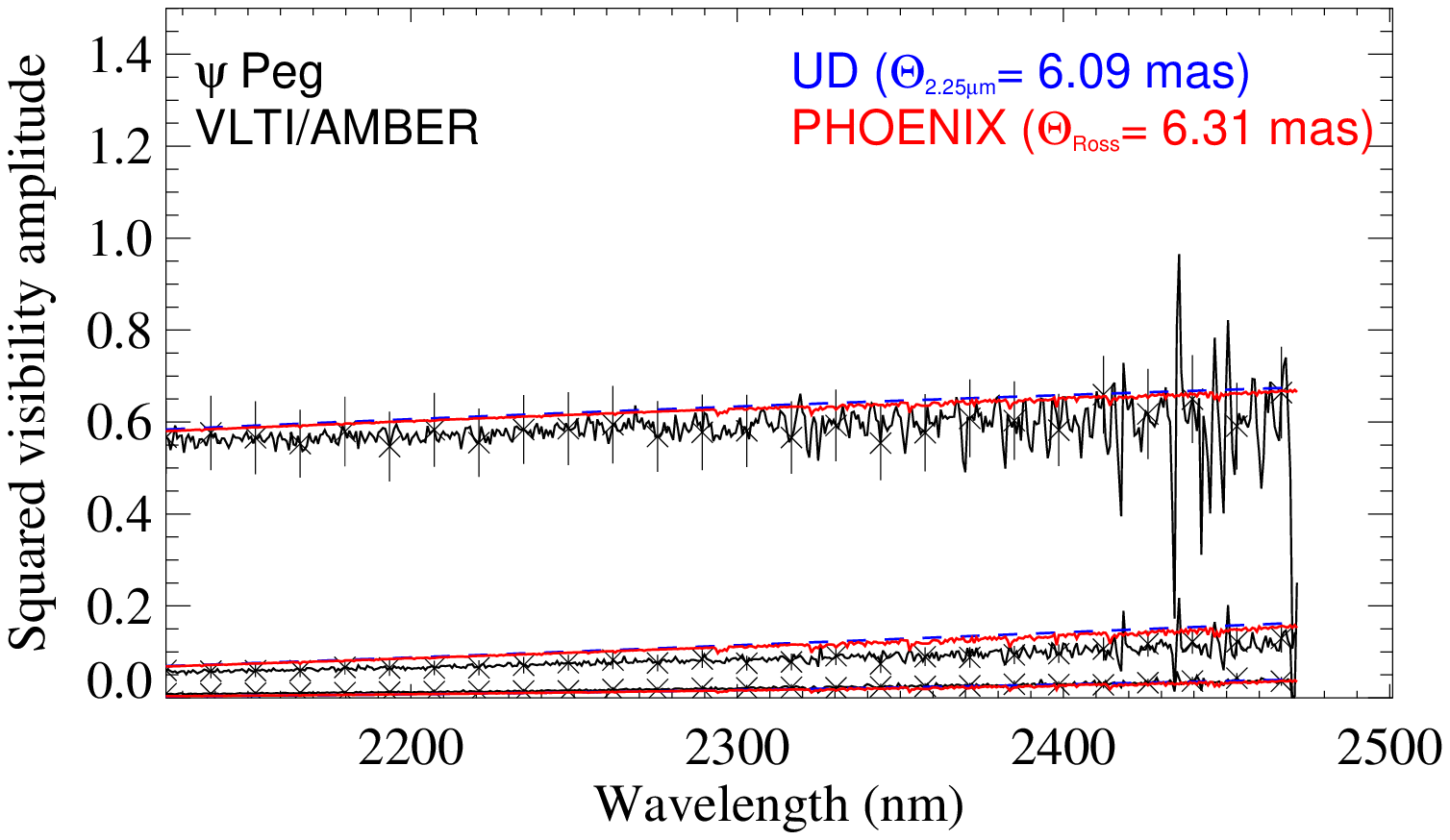}
\includegraphics[width=0.40\hsize]{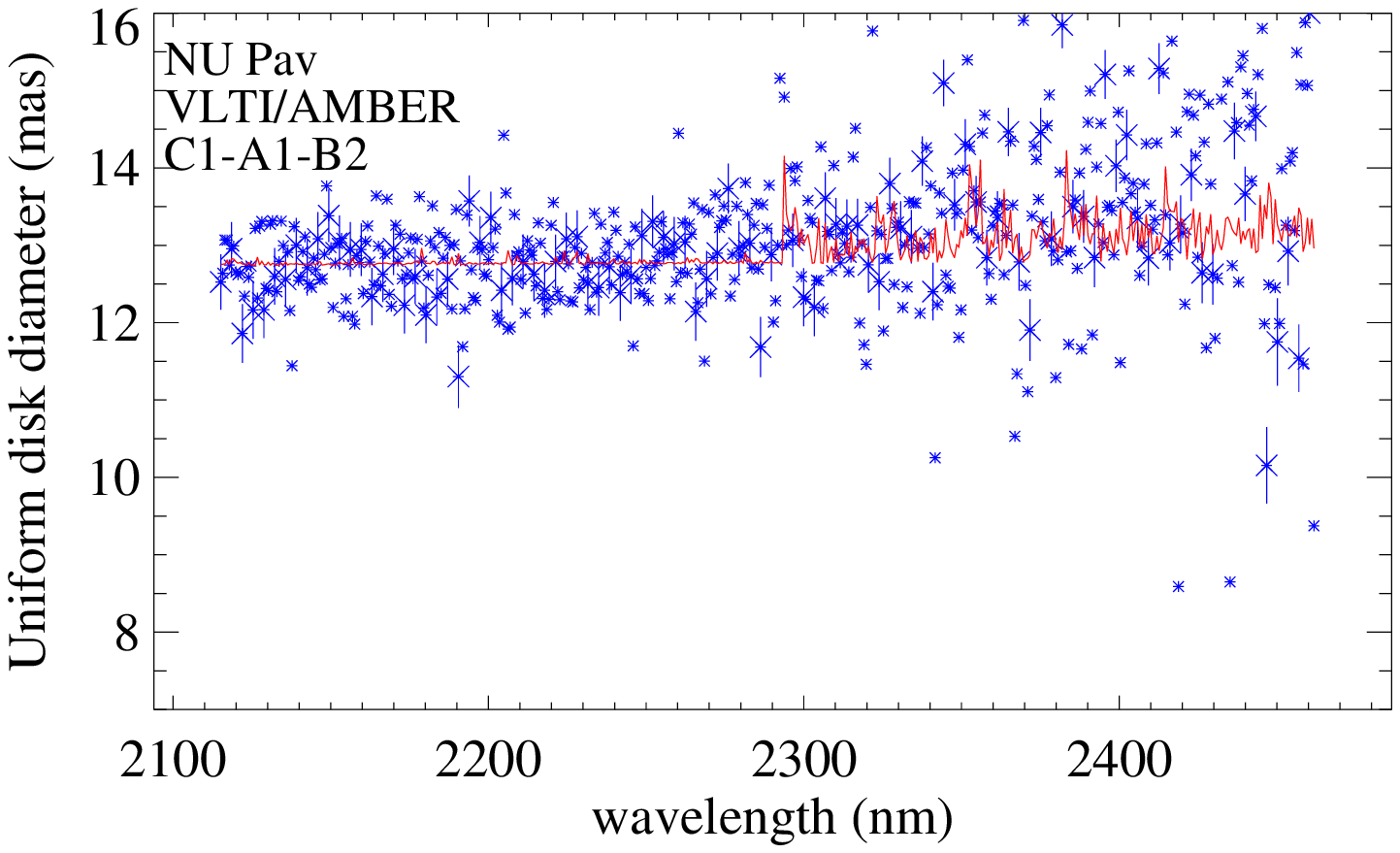}
\includegraphics[width=0.40\hsize]{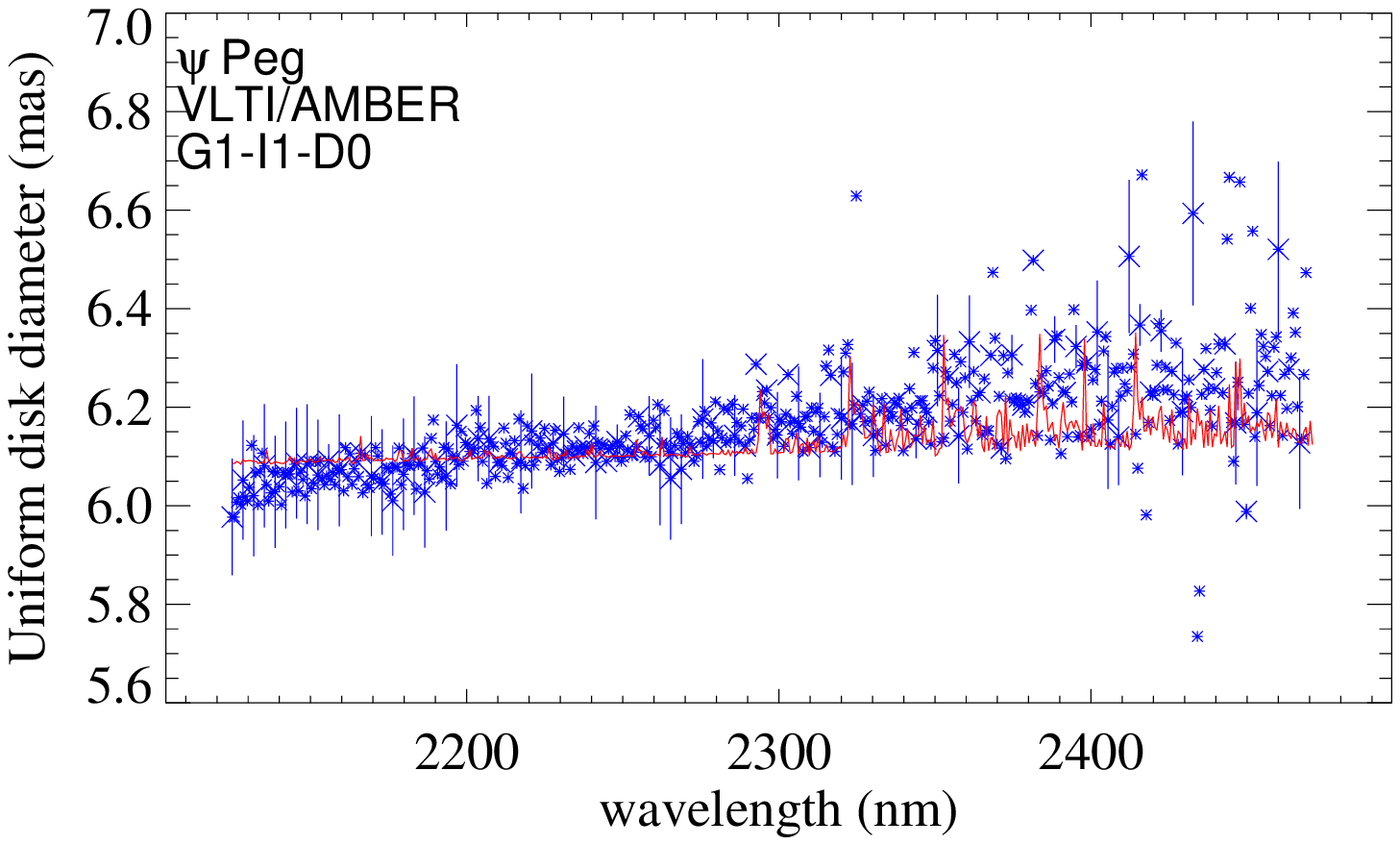}
\includegraphics[width=0.40\hsize]{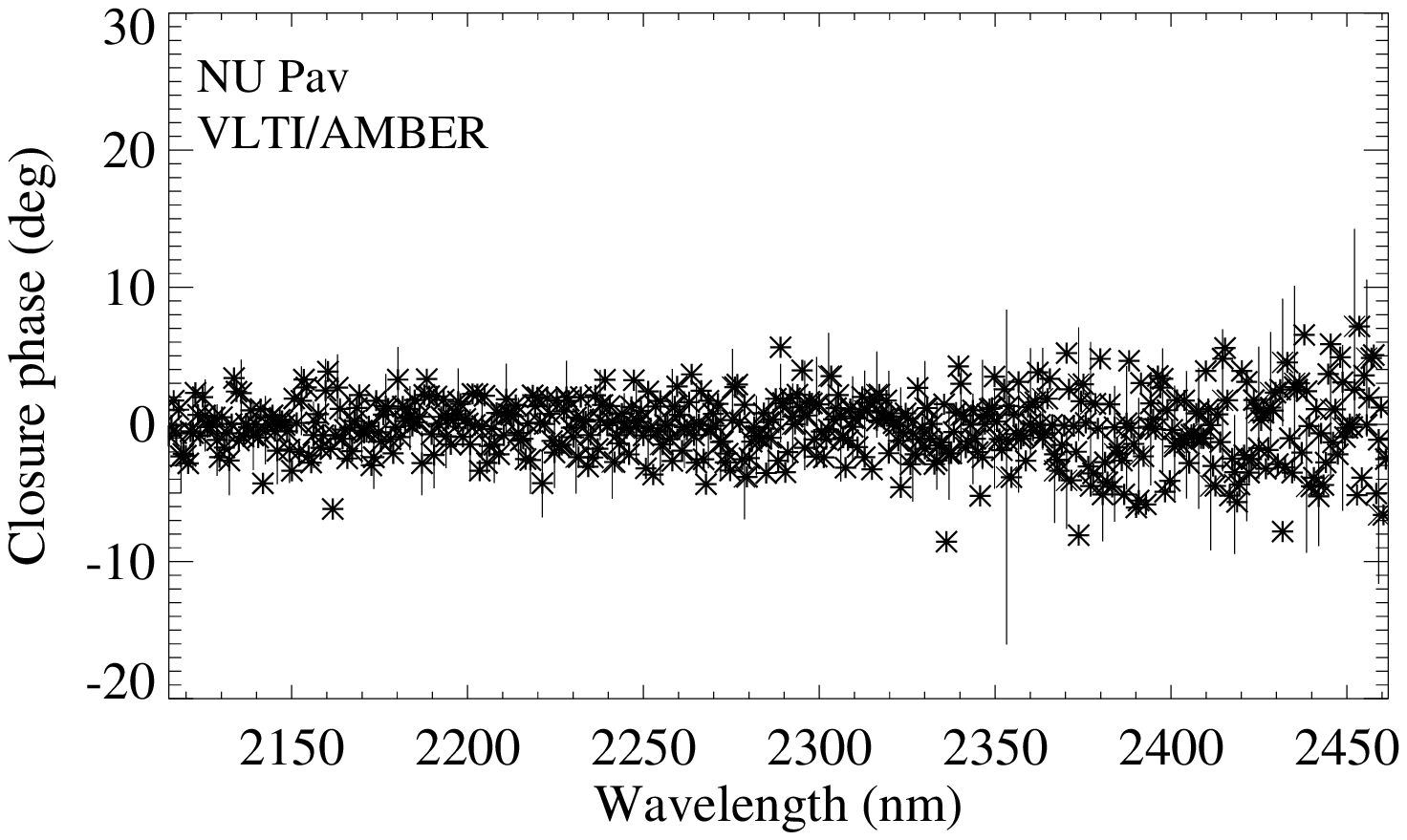}
\includegraphics[width=0.40\hsize]{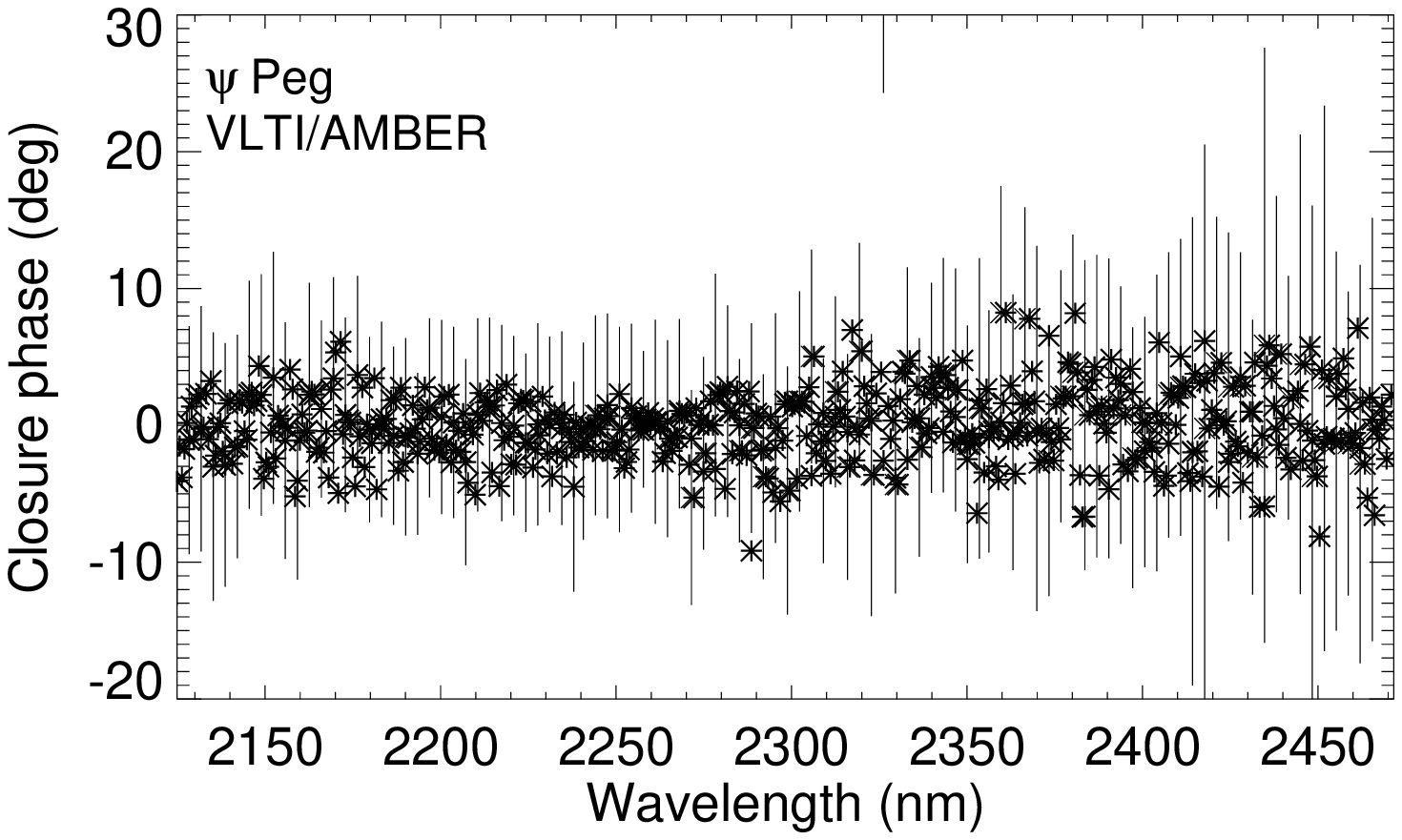}
\includegraphics[width=0.40\hsize]{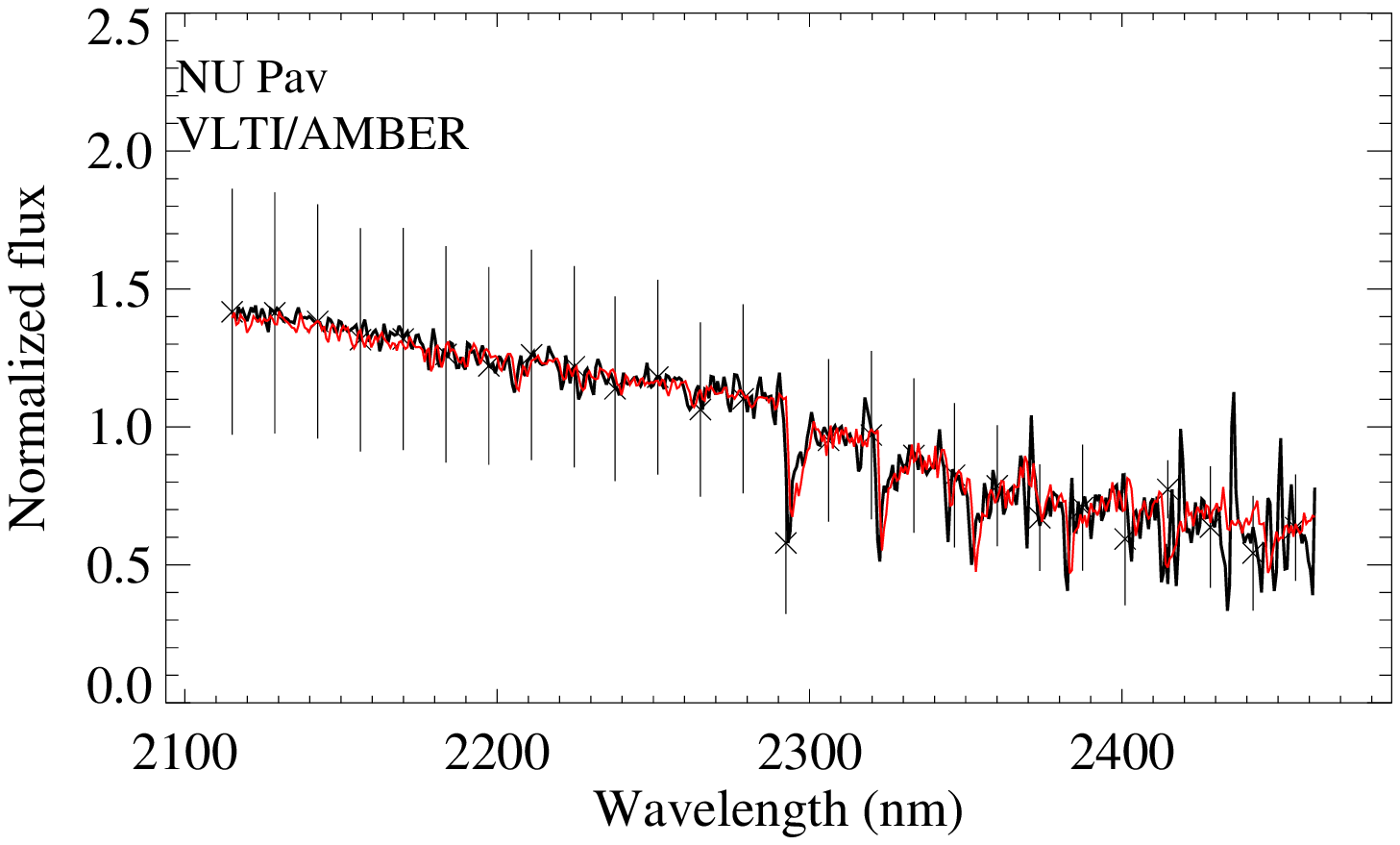}
\includegraphics[width=0.40\hsize]{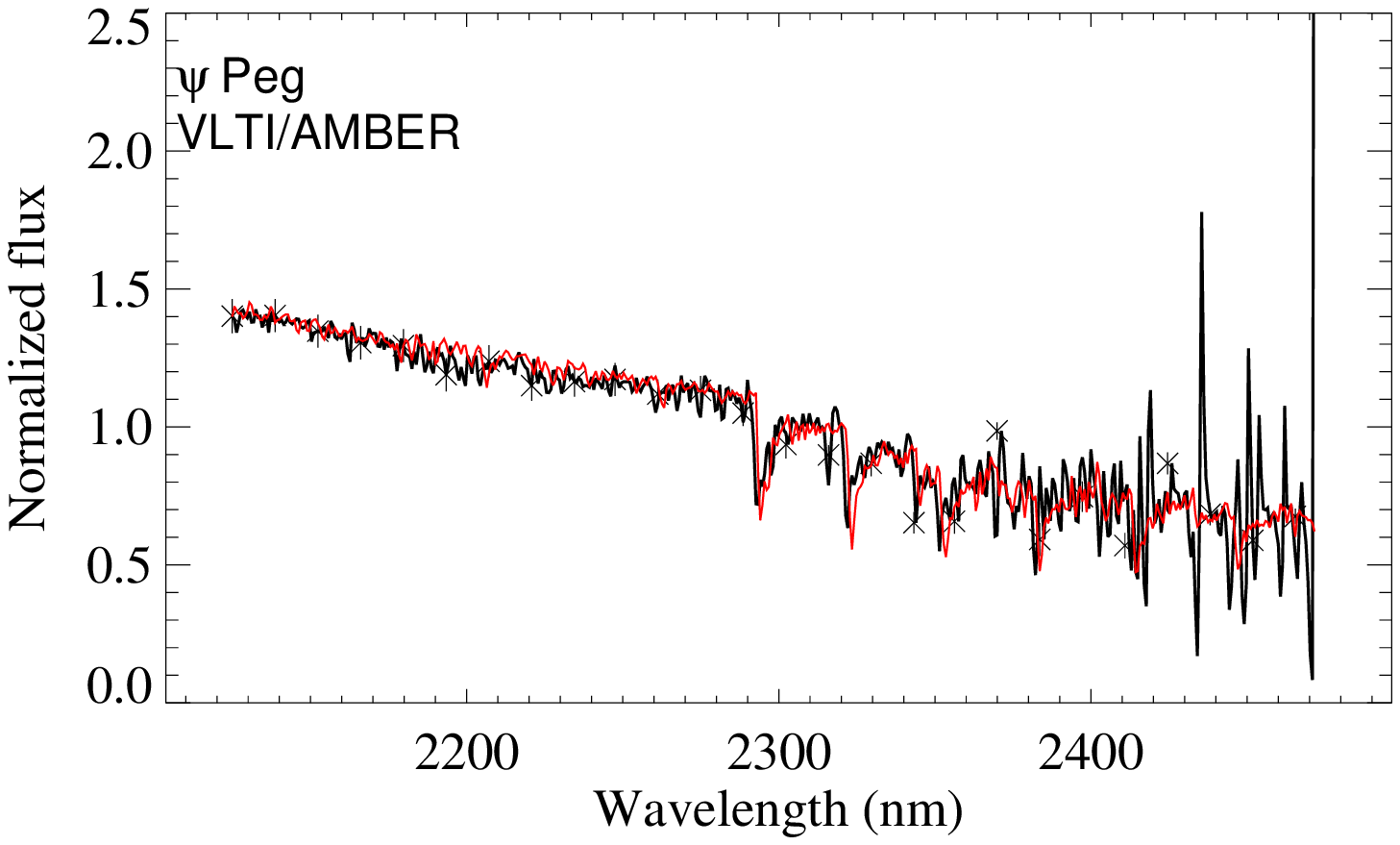}
\caption{Left: from top to bottom, observed (black) squared visibility amplitudes, UD diameters predicted from our data (blue) and from the best-fit PHOENIX model (red), closure phases in degrees, and normalized flux of NU~Pav obtained on 2012 Aug 02. Right: same as left, but for data from $\psi$~Peg obtained on 2012 Jun 16.}
\label{resul_NuPav_84Peg_fit}
\end{figure*} }

\onlfig{4}{
\begin{figure*}
\centering
\includegraphics[width=0.40\hsize]{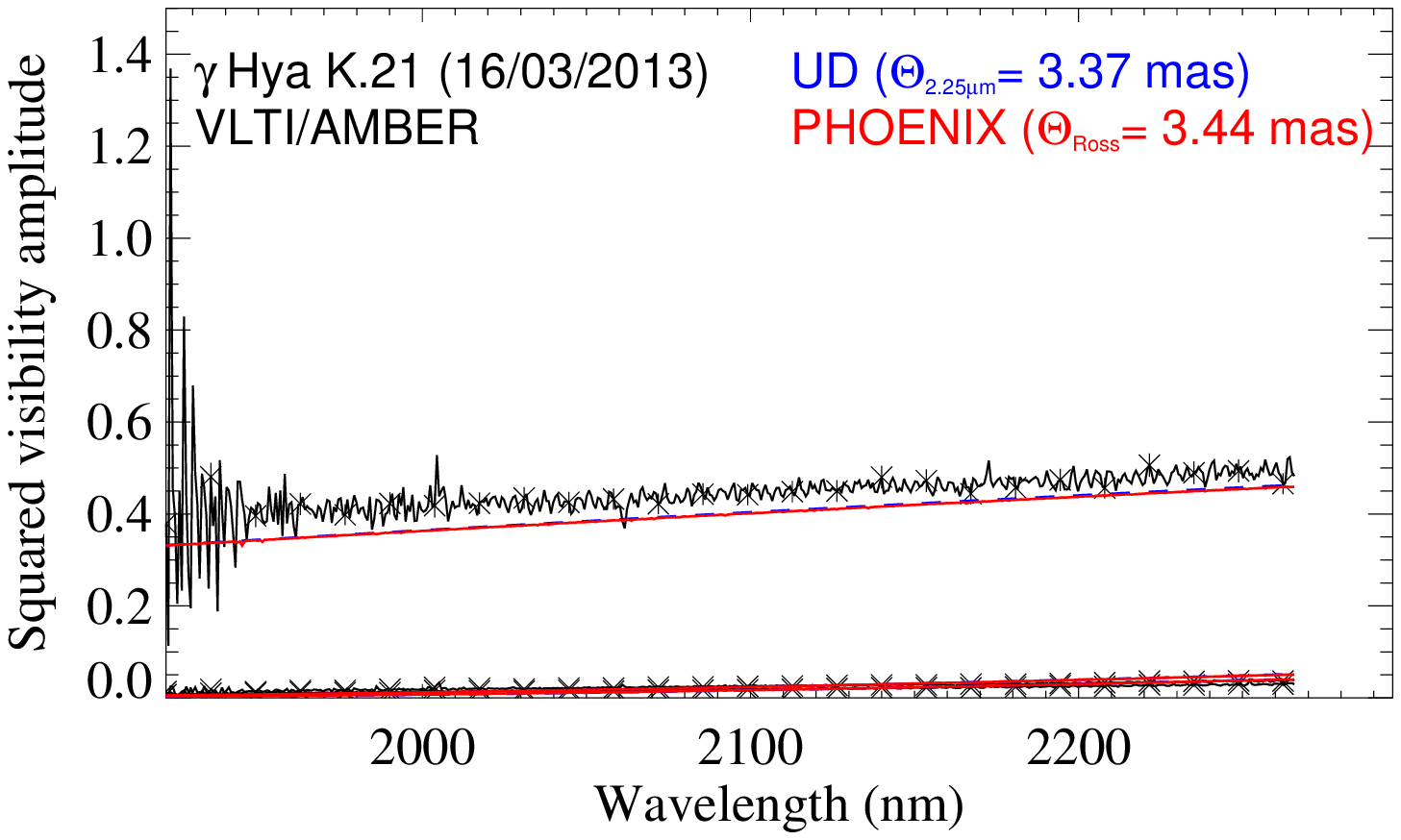}
\includegraphics[width=0.40\hsize]{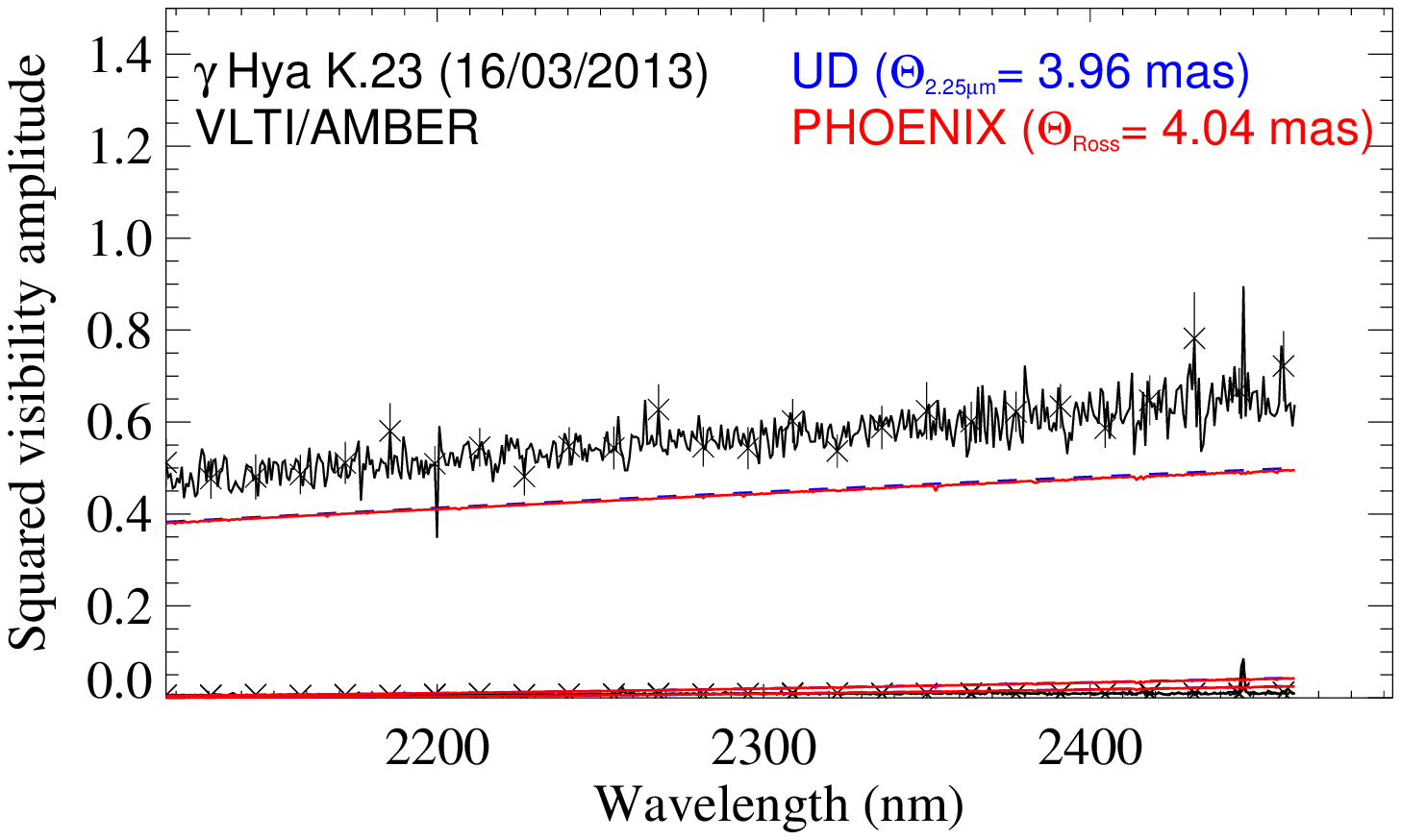}
\includegraphics[width=0.40\hsize]{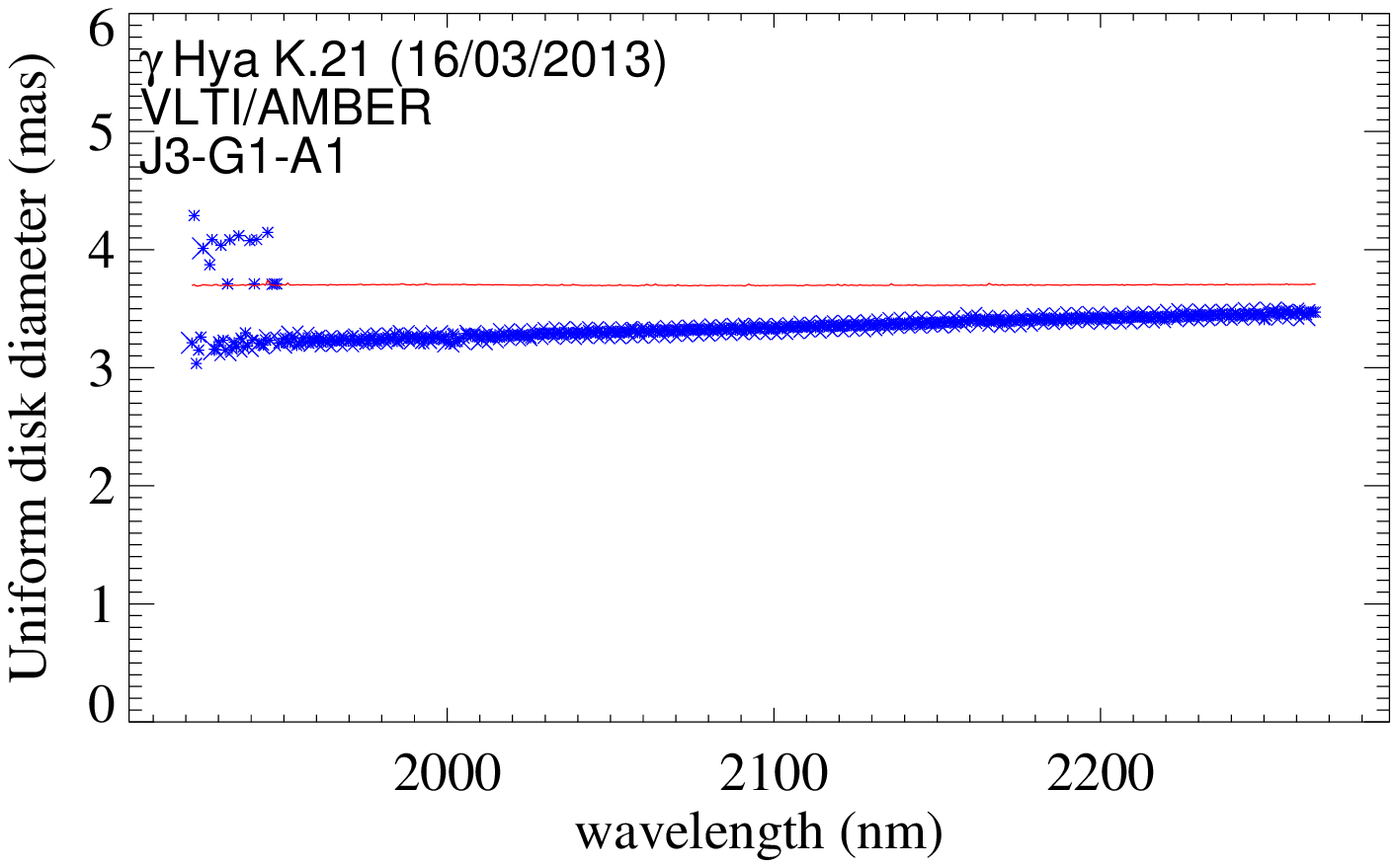}
\includegraphics[width=0.40\hsize]{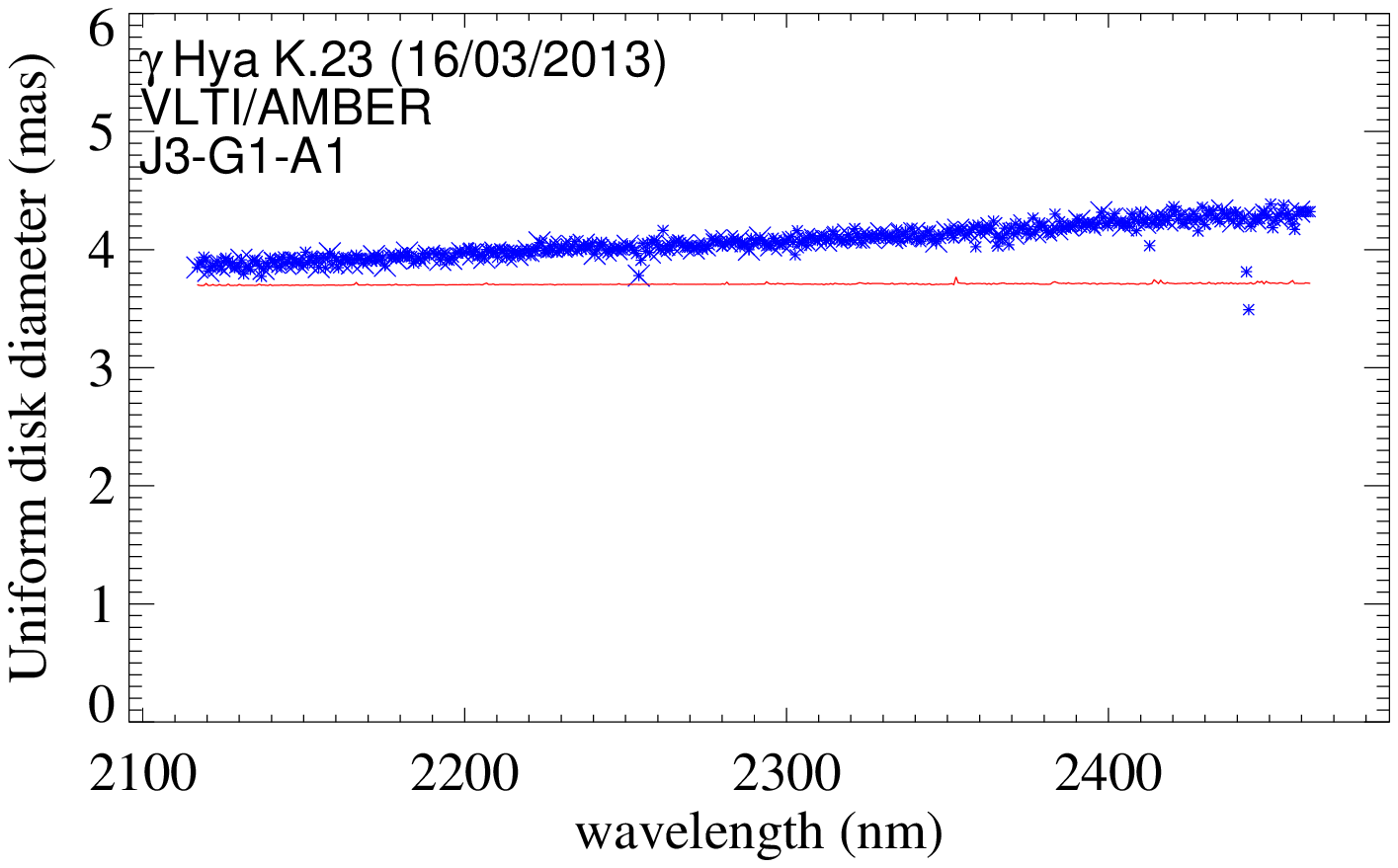}
\includegraphics[width=0.40\hsize]{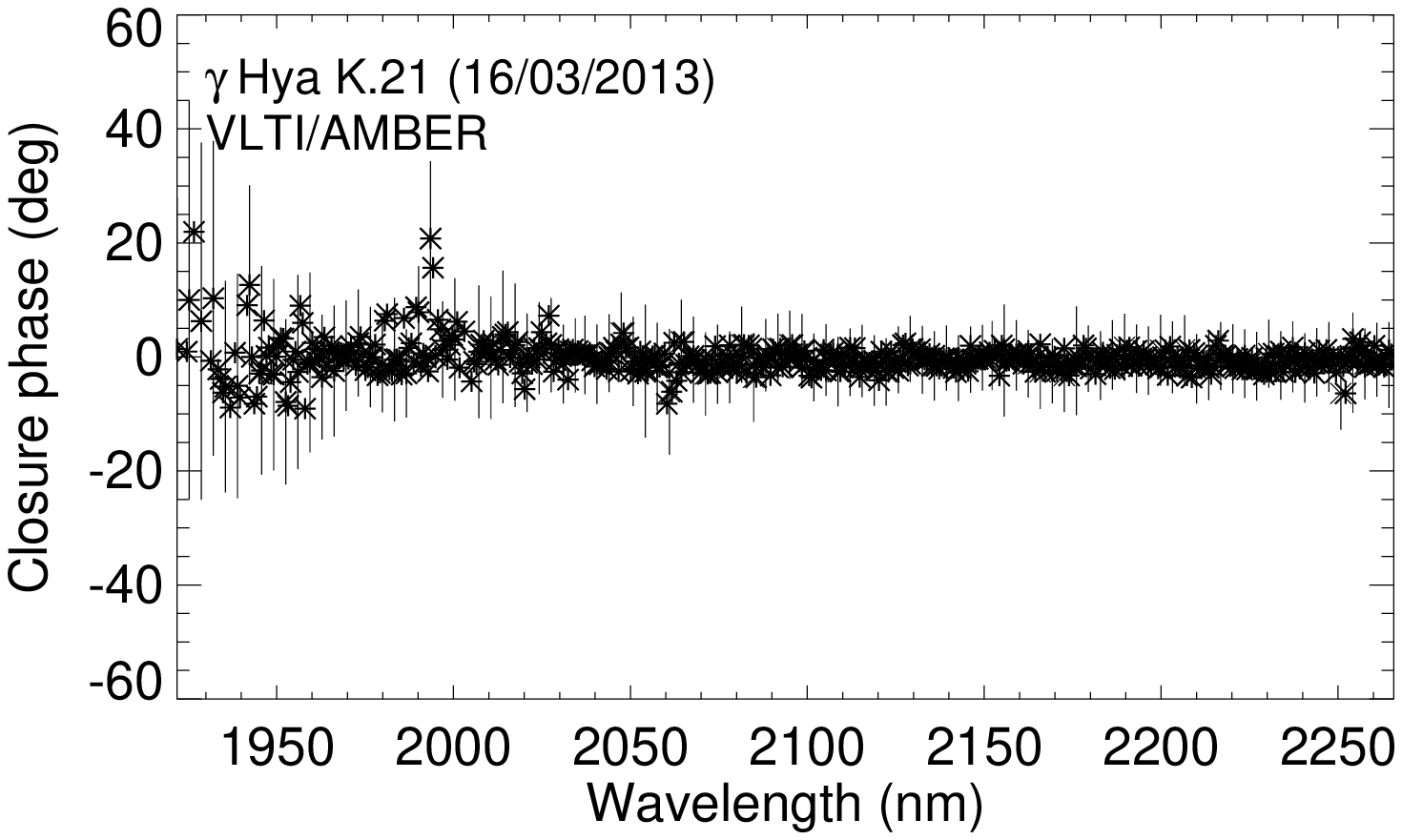}
\includegraphics[width=0.40\hsize]{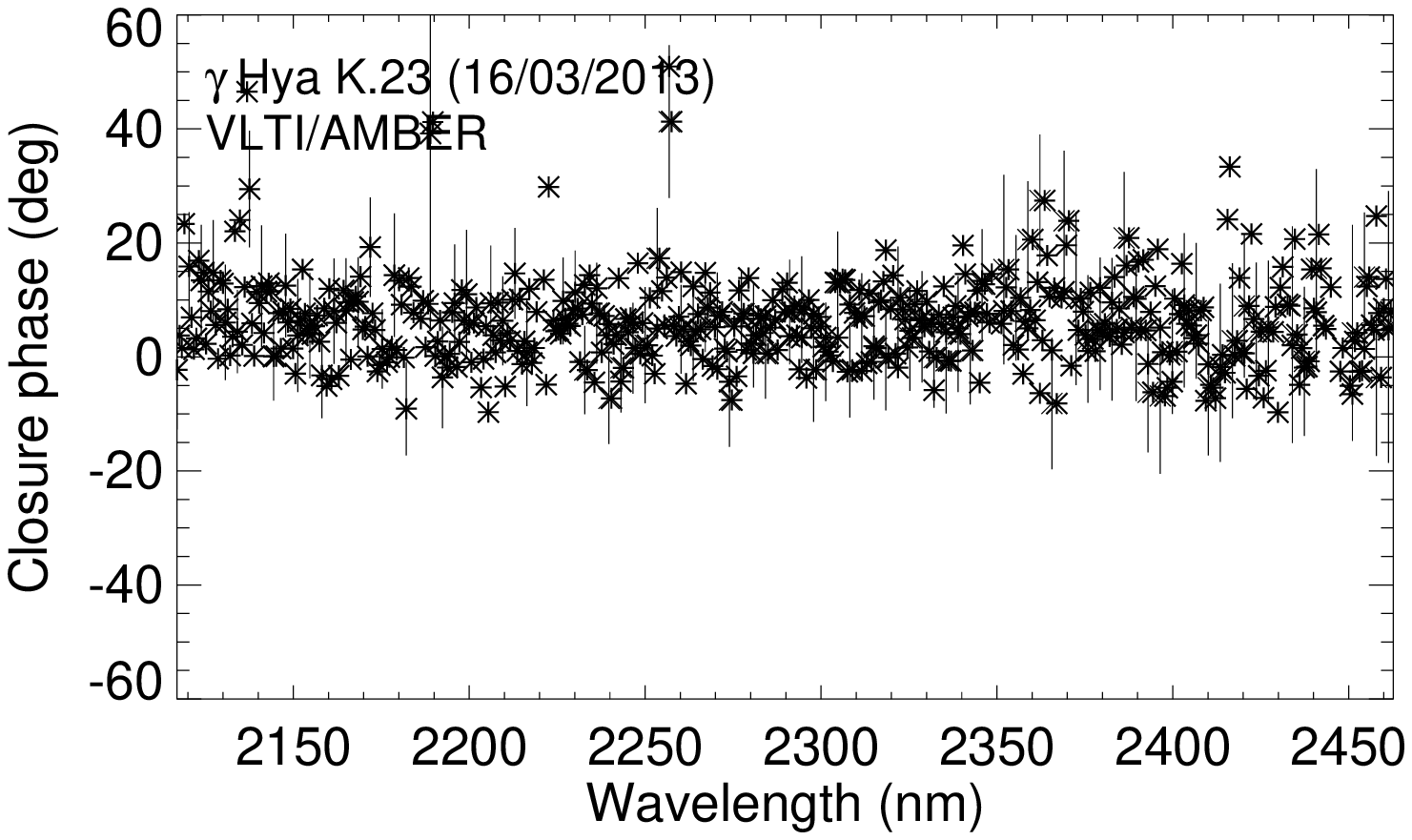}
\caption{
Left: from top to bottom, observed (black) squared visibility amplitudes, UD diameters predicted from our data (blue) and from the best-fit PHOENIX model (red), and closure phases in degrees of $\gamma$~Hya obtained on 2013 Mar 16. Right: same as left, but obtained with the MR-K 2.3\,$\mu$m setting.}
\label{resul_gammaHya_fit}
\end{figure*}}

Raw visibilities and closure phases were obtained from our AMBER data using the latest version of the \textit{amdlib} data reduction package (Tatulli et al. \cite{Tatulli2007}; Chelli et al. \cite{Chelli2009}). First, we removed the bad-pixel map and corrected for the flat contribution. Then we calculated the pixel-to-visibility matrix (P2VM) to calibrate our data for the instrumental dispersive effects, and obtained the interferometric observables. Next, we appended all scans of the same source taken consecutively and selected and averaged the resulting  visibilities of each frame using appropriate criteria. In our case, the criteria were based on the flux (we selected all frames with flux densities three times higher than the noise) and on the signal-to-noise ratio (S/N). We only used 80\% of the remaining frames with the best S/N \footnote{see AMBER Data Reduction Software User Manual; http://www.jmmc.fr/doc/approved/JMMC-MAN-2720-0001.pdf}. 

Using in-house developed IDL (Interactive Data Language) scripts we performed the absolute wavelength calibration by correlating the AMBER flux spectra  with a reference spectrum, that of the star BS~4432 (spectral type K4.5 III, similar to our calibrators; Lan\c con \& Wood \cite{Lancon2000}). A relative flux calibration of the targets was performed by using the instrumental response, estimated from the calibrators and the BS~4432 spectrum. Finally, calibrated visibility spectra were obtained by using the average of two transfer functions obtained from calibrator observations before and after each science target observation: $V^c_{sci}=V^m_{sci}/0.5(\mathcal{T}_{cal_1}+\mathcal{T}_{cal_2})$. In the case of $\epsilon$ Oct and NU~Pav, we only used one calibrator, because the visibilities of the other were not of sufficient quality. The errors of the calibrated visibilities were estimated by error propagation. For each calibrator, the error of the transfer function $\Delta \mathcal{T}$ was calculated as $\Delta \mathcal{T}=\sqrt{\Delta \mathcal{T}^2_{A}+\Delta \mathcal{T}^2_{B}+0.05^2}$. The first term was obtained as $\Delta \mathcal{T}_{A}=\Delta V^{m}_{cal}/V_{exp}$ where $\Delta V^{m}_{cal}$ were the uncertainties of the measured calibrator visibilities and $V_{exp}=2 J_1 (\pi \theta B/\lambda)/(\pi \theta B/\lambda)$, with $\theta$ the angular diameter adopted for the calibrator. The second term was defined as $\Delta \mathcal{T}_{B}=|\mathcal{T}_{cal_1}-\mathcal{T}_{cal_2}|/2$. The value of 0.05 was a systematic error adopted. This term is relevant when only one calibrator is available (then we make $\mathcal{T}_{cal_1}$=$\mathcal{T}_{cal_2}$ for practical reasons) or when by chance $\mathcal{T}_{cal_1}$ and $\mathcal{T}_{cal_2}$ are very similar.

\begin{table}
\caption{Calibration sources}
\centering
\begin{tabular}{lcccc}
\hline
\hline
 & Spectral type & Angular diameter (mas) \\
\hline
HIP 82363 & K5 III & 5.73$\pm$0.41 \\
HIP 104755 & M1.5 III & 4.20$\pm$0.30 \\
HIP 86929 & K2 II & 3.55$\pm$0.25 \\
HIP 114144 & M1 III & 4.25$\pm$0.30 \\
HIP 1168 & M2 III & 4.09$\pm$0.29 \\
K Hya & K5 III & 2.20$\pm$0.03 \\
\hline
\end{tabular}
\label{calibrator}
\end{table}


\section{Modeling: PHOENIX}

Our goals are to estimate the angular diameters of our stars and then derive their fundamental parameters. For the purpose of estimating the stellar angular diameter, we compared our observables with those provided by PHOENIX model atmospheres (version 16.03, Hauschildt \& Baron (\cite{Hausch1999}), which incorporates a model with a hydrostatic atmosphere, local thermodynamic equilibrium and spherical geometry). We used the grid corresponding to 1 Msun, that was also used by Arroyo-Torres et al. (\cite{Arroyo2013}). This grid includes effective temperatures between 2500\,K and 8000\,K in steps of 100\,K, surface gravities between log(g)=-0.5 and log(g)=4.0 in steps of 0.5 (in cgs units), and solar metallicity. Within these grids, the model provides intensity profiles for different angles of the star. To compare these models with our data, we need the flux integrated over the stellar disk and the visibility values for the baseline used. To obtain this flux, we tabulated model intensity profiles at 64 viewing angles for wavelengths 21000\,$\AA$ to 25000\,$\AA$ in steps of 0.01\,$\AA$. Then, we averaged the monochromatic intensity profiles to match the spectral channels of the individual observations and computed the data of the model for that match (see a full description of the procedure in Wittkowski et al. \cite{Witt2003}).

\subsection{Fitting the PHOENIX model to the observations}

After obtaining the grids, and using estimates of the distance and the bolometric flux (see Sect. 4.2), we proceeded to fit the PHOENIX model to our data to obtain the fundamental parameters of our stars. This iterative process was explained in Arroyo-Torres et al. (\cite{Arroyo2013}). In this case, we selected the initial T$_\mathrm{eff}$ given by Dyck et al. (\cite{Dyck1998}) and van Belle et al. (\cite{Belle2009}): T$_\mathrm{eff}$=3330\,K for $\epsilon$~Oct; T$_\mathrm{eff}$=3890\,K for $\beta$~Peg;  T$_\mathrm{eff}$=3248\,K for NU~Pav; T$_\mathrm{eff}$=3475\,K for $\psi$~Peg; and T$_\mathrm{eff}$=5087\,K for $\gamma$~Hya. We adopted an initial surface gravity of log(g)=0.0 throughout.  

After all the iterations, the final values for the PHOENIX model are for $\epsilon$~Oct, T$_\mathrm{eff}$=3500\,K, log(g)=0.0; for $\beta$~Peg, T$_\mathrm{eff}$=3800\,K, log(g)=1.0; for NU~Pav, T$_\mathrm{eff}$=3500\,K, log(g)=0.0; for $\psi$~Peg, T$_\mathrm{eff}$=3700\,K, log(g)=1.0; and for $\gamma$~Hya, T$_\mathrm{eff}$=4800\,K, log(g)=2.0. For all five cases, we used a model with solar metallicity, micro-turbulent velocity of 2\,km$/$s, and mass of 1\,$M_{\odot}$. We note that the structure of the atmosphere is not very sensitive to variations of the mass (Hauschildt et al. \cite{Hausch1999-2}). Certainly, any of those structure variations are below the level of the detectability of our interferometer.


\section{Results and discussion}

\begin{figure}
\centering
\includegraphics[width=0.85\hsize]{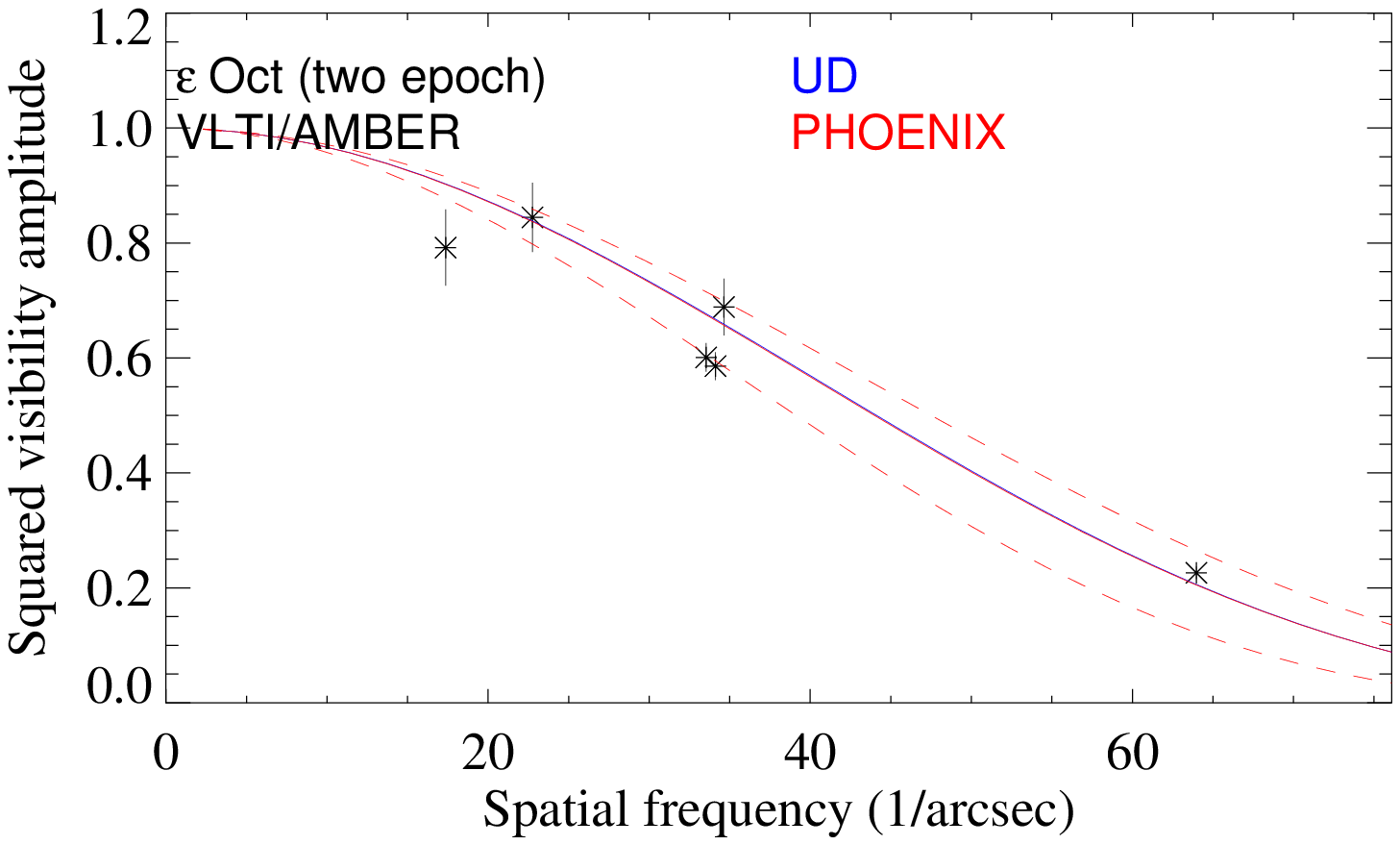}
\includegraphics[width=0.85\hsize]{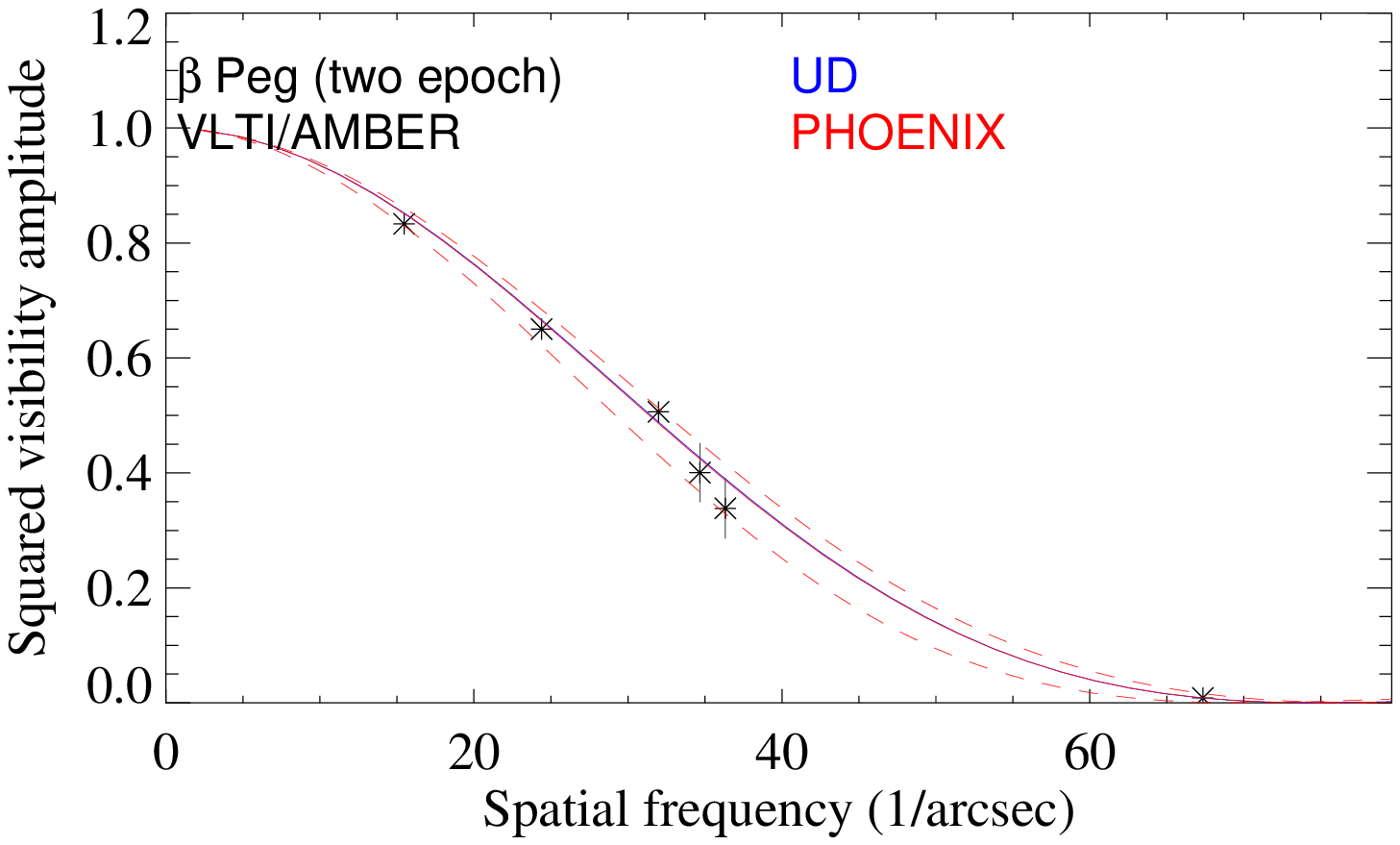}
\includegraphics[width=0.85\hsize]{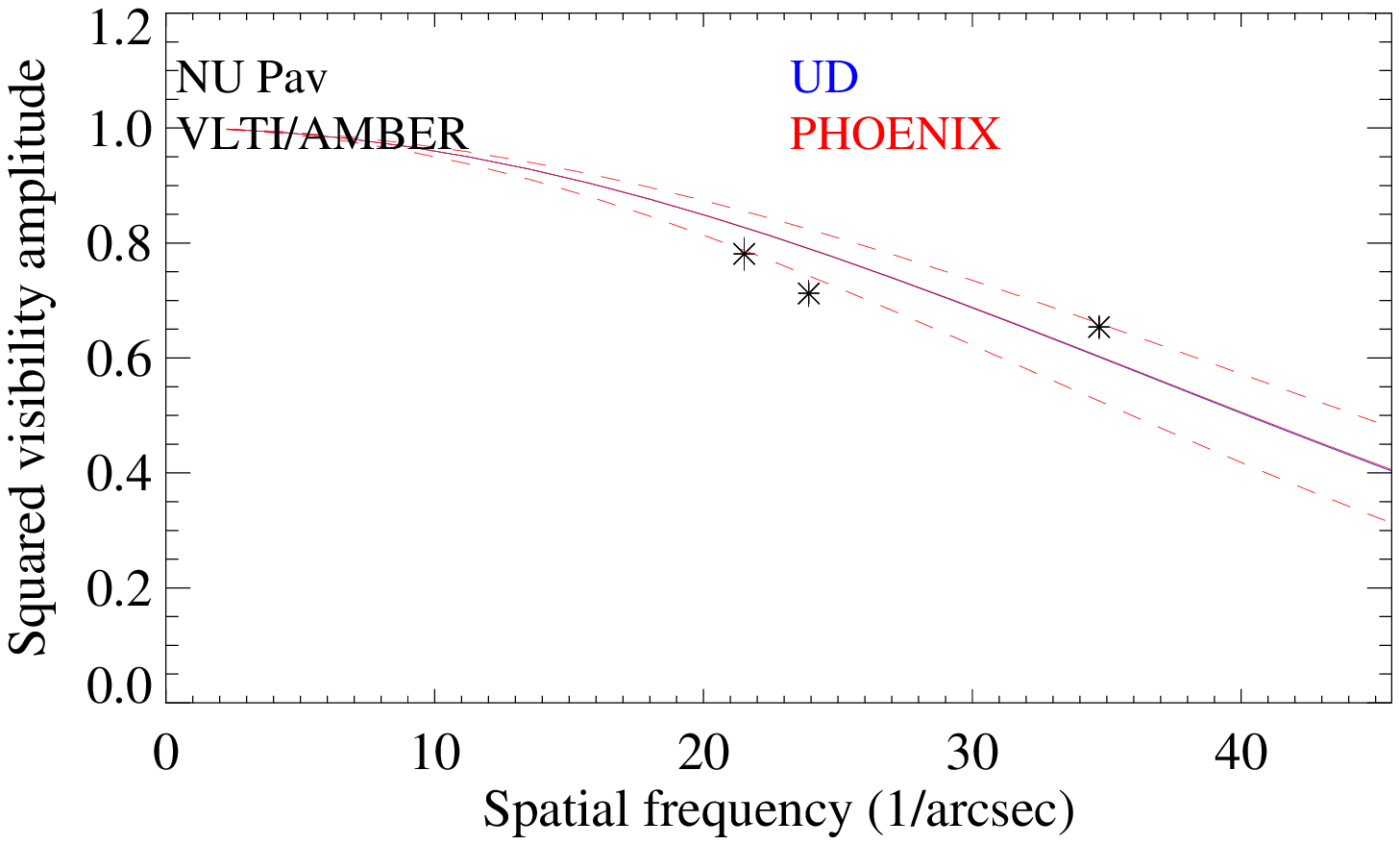}
\includegraphics[width=0.85\hsize]{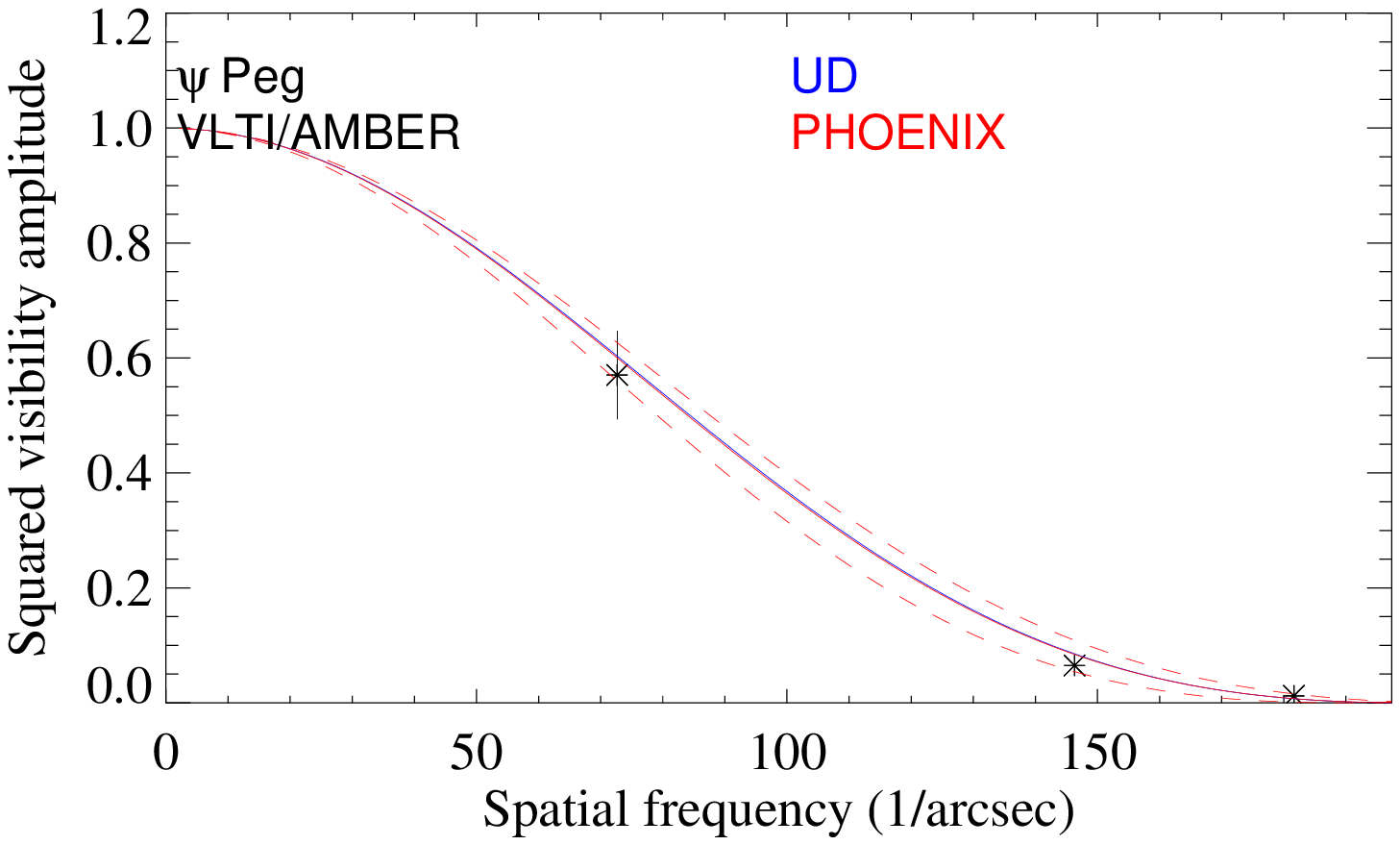}
\includegraphics[width=0.85\hsize]{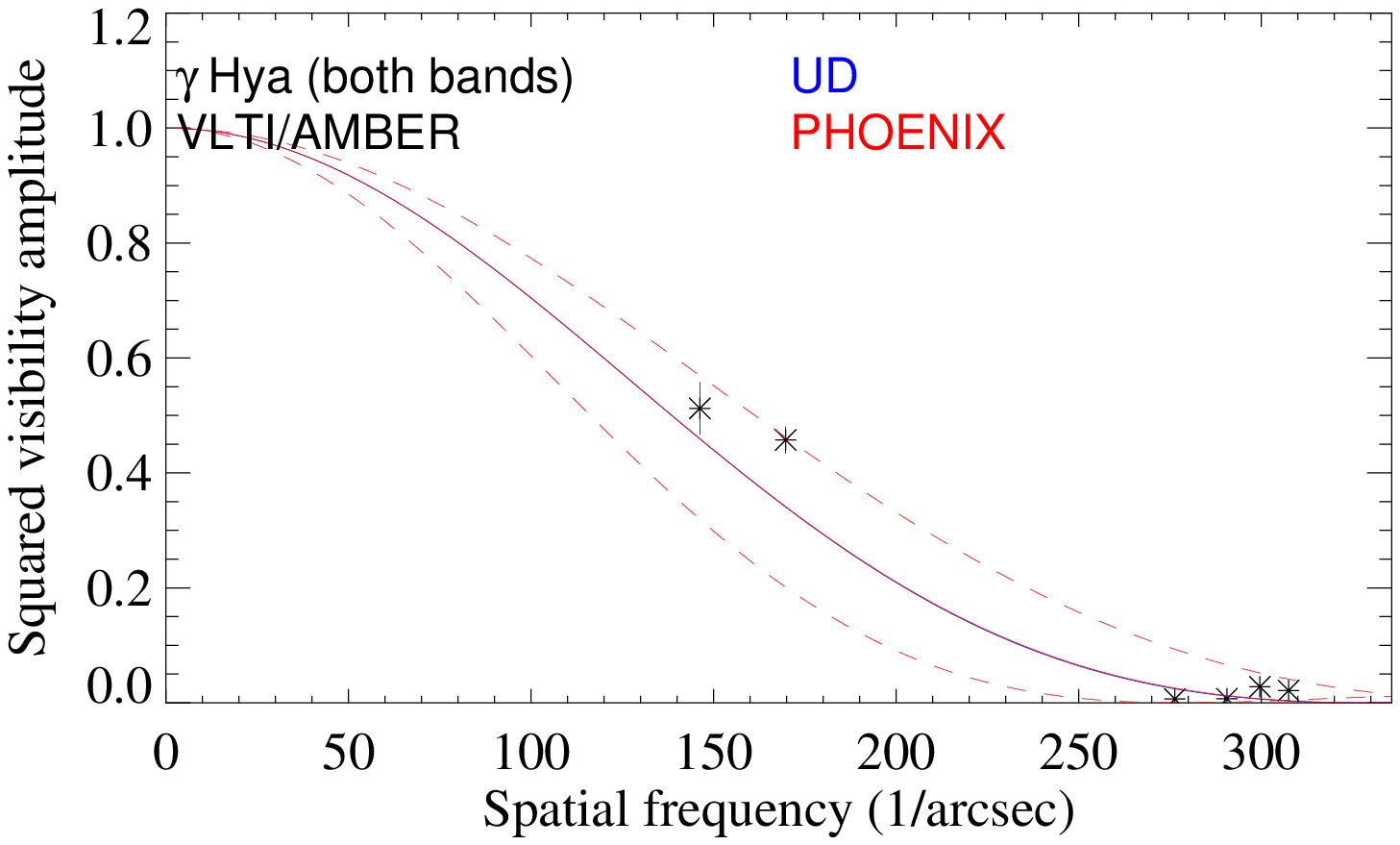}
\caption{Average (per baseline) of squared visibility amplitudes taken in the continuum bandpass
at 2.15-2.25\,$\mu$m for $\epsilon$~Oct, $\beta$~Peg, NU~Pav, $\psi$~Peg, and $\gamma$~Hya (from top to bottom) as a function of spatial frequency. For $\epsilon$~Oct and $\beta$~Peg, the graphics show the data of June and August, and for $\gamma$~Hya show the data of the both setups. The red lines indicate the best-fit UD models and the blue lines the best-fit PHOENIX models. The dashed lines are the maximum and minimum visibility curves, from which we estimated the errors of the angular diameters.}
\label{Vis_spacialFrec}
\end{figure}

Figure \ref{resul_betaPeg_fit} and Figures \ref{resul_BoOct_fit} to \ref{resul_gammaHya_fit} (only online version) show the visibility amplitudes, the closure phases, the uniform disk diameters for the stars $\epsilon$~Oct, $\beta$~Peg, NU~Pav, $\psi$~Peg, and $\gamma$~Hya, respectively, and also the fluxes for all of them but $\gamma$~Hya. For $\epsilon$~Oct and $\beta$~Peg, we have observations of two different epochs (2012 June -left panel- and 2012 August - right panel), while for $\psi$~Peg, NU Pav, and $\gamma$~Hya, we only have observations of one epoch (2012 June, 2012 August, 2013 March, respectively).  

The visibility curves are shown in the top panels of Fig. \ref{resul_betaPeg_fit} and Figs. \ref{resul_BoOct_fit} to \ref{resul_gammaHya_fit} (only online version). In these figures, we see no visibility decrease at the positions of the CO lines (in contrast with the results shown in Mart{\'{\i}}-Vidal et al. \cite{Marti2011}, for RS~Cap, or in Arroyo-Torres et al. \cite{Arroyo2013}, for a sample of RSGs). Similarly, these is no drop in the visibility between 2.3\,$\mu$m and 2.5\,$\mu$m either. For our sources (except for $\beta$~Peg), the synthetic visibilities are consistent with our observations. Since the PHOENIX model uses a hydrostatic atmosphere model and a limb-darkened disk, $\epsilon$~Oct, NU~Pav, $\psi$~Peg, and $\gamma$~Hya are compatible with a limb-darkened disk and a hydrostatic atmosphere. On the other hand, $\beta$~Peg shows a decrease of the visibility in the CO lines greater than the synthetic visibilities, and also a weak systematic trend in the range 2.3-2.5\,$\mu$m, like that observed in RS Cap.

\begin{table}[h]
\caption{Information about the variability}
\centering
\begin{tabular}{lcccccc}
\hline
\hline
 & $\epsilon$ Oct & $\beta$ Peg & NU Pav & $\psi$~Peg  & $\gamma$ Hya & Ref.\\
 \hline
Period (d) & 53 & 43.3 & 60 & -  & - & 1\\
V$_{max}$ & 4.58 & 2.31 & 4.91 & 4.63 & 2.94 & 1\\
V$_{min}$ & 5.3 & 2.74 & 5.26 & 4.69 & 3.02 & 1\\
$\Delta$V & 0.72 & 0.43 & 0.35 & 0.06 & 0.11 & - \\
Type & SRB & LB & SRB & giant & variable & 2\\
\hline
\end{tabular}
\tablefoot{1: Watson et al. (\cite{Watson2006}) 2: Samus et al. (\cite{Samus2007}) \footnotemark. The variation of the magnitude is $\Delta$V=V$_{max}$-V$_{min}$.}
\label{period}
\end{table} 
\footnotetext{http://www.sai.msu.su/gcvs/gcvs/}

The uniform disk diameters calculated from our data as a function of wavelength are shown in the second panel of Fig. \ref{resul_betaPeg_fit} and Figs. \ref{resul_BoOct_fit} to \ref{resul_gammaHya_fit} (only online version). In the observations from 2012 August, we see a larger scatter in the data, which prevents us from drawing any firm conclusion about the possible larger diameter in the CO lines (the atmospheric conditions were poor these nights). In the case of $\epsilon$~Oct, the data from 2012 Jun 25 show that the uniform disk diameter is constant across the band. Thus, the CO bandheads do not present a larger size than for the continuum of the star. The synthetic visibilities and our observations are consistent. The $\beta$~Peg data observed in 2012 Jun 25 show a size in the CO band similar to that predicted by the PHOENIX model, but this source also presents an additional slope in the data that is not predicted by the PHOENIX model. This slope is perhaps due to a layer of H$_{2}$O that is not present in the model (as in RS~Cap, Mart{\'{\i}}-Vidal et al. \cite{Marti2011}). The increase of the angular diameter in the CO region is about 5.3\% with respect to the near-continuum bandpass diameter. The NU~Pav data have so much scatter that we cannot discern any change in the angular diameter of the region related to the CO lines. The $\psi$~Peg data show a very small increase of the angular diameter in the CO band, similar to the one synthesized in the PHOENIX model. Finally, the uniform disk diameter for $\gamma$~Hya is constant in both bands (K-2.1 and K-2.3), with the size at the the CO bandheads not larger than in the continuum. In summary, the PHOENIX predictions agree well with the data, showing that the simulated atmospheres are as compact as the observed ones, except for $\beta$~Peg, with a 5.3\% size increase clearly visible in the data but not modeled by PHOENIX.

The closure phases are shown in the third panel of Fig. \ref{resul_betaPeg_fit} and Figs. \ref{resul_BoOct_fit} to \ref{resul_gammaHya_fit} (only online version). They show low values ($\leq$\,20\,deg) across the band, indicating little or no deviations from point symmetry. However, as our measurements lie in the first visibility lobe, we cannot exclude asymmetries on scales smaller than the observed stellar disk. 

The fourth panel in Fig. \ref{resul_betaPeg_fit} and Figs. \ref{resul_BoOct_fit} to \ref{resul_NuPav_84Peg_fit} (only online version) shows the normalized flux spectra of our targets. For $\gamma$~Hya we do not have spectra data because there was an error with the data formatting during the observation. In the normalized spectra, we observe a decreasing flux from 2.25\,$\mu$m and strong absorption lines of CO. The synthetic spectra of the PHOENIX model agree well with our data, in particular in the CO bandheads. This indicates that the opacities of CO lines in cool giant stars are well reproduced by the PHOENIX model (as found previously by Lan\c con et al. \cite{Lancon2007}).

We note that the sizes estimated in the CO bandheads of all the stars in our sample match the extension of the CO layers predicted by the PHOENIX hydrostatic models (with the exception of $\beta$\,Peg, for which an additional layer of water vapor may be necessary). In contrast, much more extended CO layers, which cannot be reproduced by hydrostatic models, have been reported in other AGB and RSG stars (Mart{\'{\i}}-Vidal et al. \cite{Marti2011}; Wittkowski et al. \cite{Wittkowski2008}, \cite{Witt2012};  Arroyo-Torres et al. \cite{Arroyo2013}).

The stars reported here are semiregular late-type giants ($\epsilon$~Oct and NU~Pav), an irregular pulsating variable ($\beta$~Peg), a normal giant ($\psi$~Peg), and a low-amplitude variable ($\gamma$~Hya). These stars are characterized by short and irregular periods, with a low or very low variability amplitude ($\Delta$V between 0.06 and 0.72; see Table \ref{period}). However, the stars reported in Arroyo-Torres et al. (\cite{Arroyo2013}) and Mart{\'{\i}}-Vidal et al. (\cite{Marti2011}) are semiregular giants with much longer periods and variability amplitudes ($\Delta$V between 2.0 and 2.3).

The larger extension of the CO layers in stars with higher variability may be indicative of the important role of strong pulsations for the mechanical transport of the molecular gas, away from the stellar surface. These pulsations trigger the onset of the stellar wind (by radiative pressure on the condensed dust grains) according to the standard model of strong winds in evolved giant stars.

\subsection{Estimate of the angular diameter}

After the best PHOENIX models were fitted to our data, we estimated the angular diameter for each source. To estimate the angular diameter, we used the continuum interferometric data near 2.20\,$\mu$m, free of contamination from the CO band and well reproduced by a limb-darkened disk. 

The angular diameter as obtained from the best-fit to the PHOENIX model corresponds to the size of the outermost layer (0\% intensity). To estimate the Rosseland angular diameter (corresponding to the layer where the Rosseland optical depth equals 2/3),  we multiplied our value of the angular diameter by the ratio between the outermost layer and the Rosseland layer. This ratio was 0.79 (for $\epsilon$~Oct and NU~Pav), 0.93 (for $\beta$~Peg and $\psi$~Peg), and 0.97 (for $\gamma$~Hya). The resulting Rosseland angular diameters and the angular diameters obtained from the UD model are shown in Table \ref{angular_diam}. Results for $\epsilon$~Oct and $\beta$~Peg were obtained using the two available epochs (2012 June and 2012 August), and results for $\gamma$~Hya were estimated using both setups (K-2.1 and K-2.3). The errors include statistical and systematic errors caused by calibration uncertainties. We conservatively estimate these errors from the differences between the visibility curves lying at the maximum and minimum of our data.

The angular diameter of $\beta$~Peg, $\psi$~Peg, and $\gamma$~Hya were estimated previously by Richichi et al. (\cite{Richichi2005}). Their angular diameter estimates were $\theta _{UD}$=16.19$\pm$0.23\,mas and $\theta _{LD}$=16.75$\pm$0.24\,mas for $\beta$~Peg, $\theta _{UD}$=6.40$\pm$0.60\,mas for $\psi$~Peg, $\theta _{UD}$=2.96$\pm$0.15\,mas for $\gamma$~Hya. These values are compatible with our estimates shown in Table \ref{angular_diam}.

\begin{table*}
\caption{Summary of estimated angular diameters}
\centering
\begin{tabular}{lccccc}
\hline
\hline
  & $\epsilon$ Oct & $\beta$ Peg & NU Pav & $\psi$ Peg & $\gamma$ Hya\\
\hline
$\theta _\mathrm{Ross}$  & 11.66$\pm$1.50 mas & 16.87$\pm$1.0 mas & 13.03$\pm$1.75 mas & 6.31$\pm$0.35 mas & 3.78$\pm$0.65 mas \\ 
$\theta _\mathrm{UD}$  & 11.42$\pm$1.50 mas & 16.32$\pm$1.0 mas & 12.78$\pm$1.75 mas & 6.09$\pm$0.35 mas & 3.71$\pm$0.65 mas \\
\hline
\end{tabular}
\label{angular_diam}
\end{table*} 

Figure \ref{Vis_spacialFrec} shows averaged continuum visibility data as a function of spatial frequency and the visibility curves corresponding to the extreme value covered by our uncertainty in the angular diameter. The visibility errors shown in this figure were estimated as an average of individual errors, since the errors are dominated by systematic effects. The model fit itself was made using all the individual data points, not those averaged. According to the results of Figure \ref{Vis_spacialFrec}, our PHOENIX estimated angular diameter is compatible with our observations. In $\beta$~Peg, $\psi$~Peg and $\gamma$~Hya, we have several points sampling up to the first visibility null. Knowing the position of the null gives higher precision to the estimate of the angular diameter.  

\begin{table*}
\caption{Fundamental parameters of $\epsilon$~Oct, $\beta$~Peg, NU~Pav, $\psi$~Peg, and $\gamma$~Hya.}
\begin{center}
\begin{tabular}{ccccccc}
\hline
\hline
Parameter  & $\epsilon$ Oct & $\beta$ Peg &  NU Pav & $\psi$ Peg & $\gamma$~Hya & Ref.   \\
\hline
$E_{B-V}$ (mag)  & 0.505 & 0.294 & 0.476 & 0.13 & 0.18 & this work\\
$F_{bol}$   & 7.27$\pm$1.09 & 22.10$\pm$3.32 &  8.65$\pm$1.30 & 2.50$\pm$0.38 & 2.39$\pm$0.36 & 1  \\ 
($10^{-9}$ W $m^{-2}$) & & & & \\
d (pc)  & 89.08$\pm$1.78 & 60.09$\pm$0.54 &  145.52$\pm$5.53 & 145.44$\pm$5.10 & 41.03$\pm$0.25 & 2 \\
L ($10^{29}$ W)  & 6.91$\pm$1.07 & 9.56$\pm$1.44 &  21.90$\pm$3.68 & 6.32$\pm$1.05 & 0.48$\pm$0.07 &  1,2 \\
log(L/$L_{\odot}$)  & 3.26$\pm$0.16 & 3.40$\pm$0.15 & 3.76$\pm$0.17 & 3.22$\pm$0.17 & 2.10$\pm$0.15 & - \\
$\theta _{Ross}$ (mas)  & 11.66$\pm$1.50 & 16.87$\pm$1.00 & 13.03$\pm$1.75 & 6.31$\pm$0.35 & 3.78$\pm$0.65 & this work \\
R($R_\odot$)  & 112$\pm$15 & 109$\pm$7 & 204$\pm$29 & 98$\pm$6 & 16$\pm$3 & 2, this work \\
$T_\mathrm{eff}$ (K)  & 3560$\pm$264 & 3909$\pm$187 & 3516$\pm$275 & 3705$\pm$177 & 4727$\pm$444  & 1,2, this work \\
log($T_\mathrm{eff}$)  & 3.55$\pm$0.07 & 3.59$\pm$0.05 & 3.55$\pm$0.08 & 3.57$\pm$0.05 & 3.67$\pm$0.09 & -  \\ 
log(g) & 0.3 & 0.8 & 0.2 & 0.6 & 2.2 & this work \\
\hline
\end{tabular}
\tablefoot{1: Kharchenko (\cite{Kharchenko2001}), Morel \& Magnenat (\cite{Morel1978}), Cutri et al. (\cite{2mass}), IRAS (\cite{iras}). 2: Anderson \& Francis (\cite{Anderson2012}). We assumed a 15\% error in the flux. The distance error was based on the values from Anderson \& Francis (\cite{Anderson2012}). The errors in the luminosity, effective temperature, and radius were estimated from error propagation.}
\end{center}
\label{fund_parameters}
\end{table*}

\subsection{Fundamental parameters}

Our angular diameter results allow us to obtain estimates of fundamental parameters of the observed stars, namely, the effective temperature, the radius, and the luminosity. The angular diameters are those calculated in Sect. 4.1 and shown in Table \ref{angular_diam}. The effective temperature is based on the angular diameter and the bolometric flux; the radius is estimated from the angular diameter and the adopted distance; and the luminosity is derived from the bolometric flux and the distance. We used the distance values from Anderson \& Francis (\cite{Anderson2012}).

For the bolometric flux of our targets, we used the BVRIJHKL magnitudes from Kharchenko (\cite{Kharchenko2001}), Morel \& Magnenat (\cite{Morel1978}), and Cutri et al. (\cite{2mass}). We also used the IRAS flux from IRAS (\cite{iras}). To convert the magnitudes into fluxes, we used the zero values from Skinner (\cite{Skinner1996}) and the 2MASS (Cohen et al. \cite{Cohen2003}) system. To deredden the flux values we used the color excess method applied to (V-K), because photometric colors with longer spectral baselines provide more accurate results. The color excess $E_{V-K}$ is calculated as the difference between the observed color $(V-K)_{star}$ of our star and the intrinsic color $(V-K)_{0}$, obtained from Ducati et al. \cite{Ducati2001} as a function of the spectral type of the star. We converted the $E_{V-K}$ into $E_{B-V}$ using the following equations for M-type stars given by Fiorucci \& Munari (\cite{Fiorucci2003}):

$$\frac{A_K}{A_V}=0.12$$ 
$$\frac{A_V}{E_{B-V}}=3.69 \: \mathrm{,and}$$ 
$$E_{B-V}=\frac{1}{(1-A_K/A_V)\ast 3.69}\ast E_{V-K}.$$

The obtained $E_{V-K}$ values are summarized in Table \ref{fund_parameters}. After that, we calculated the absorption in all photometric bands ($A_\lambda$) by means of the equation $A_\lambda =(\alpha _{\lambda} + \beta _{\lambda}\ast E_{B-V})\ast E_{B-V}$, where the values of $\alpha _{\lambda}$ and $\beta _{\lambda}$ are also taken from Fiorucci \& Munari (\cite{Fiorucci2003}). Finally, we corrected all the flux values for interstellar extinction and integrated them to obtain the bolometric flux (we used the Newton-Cotes integration method within the IDL program). Table \ref{fund_parameters} summarizes the fundamental parameters for our targets. 

\begin{figure}
\centering
\includegraphics[width=0.85\hsize]{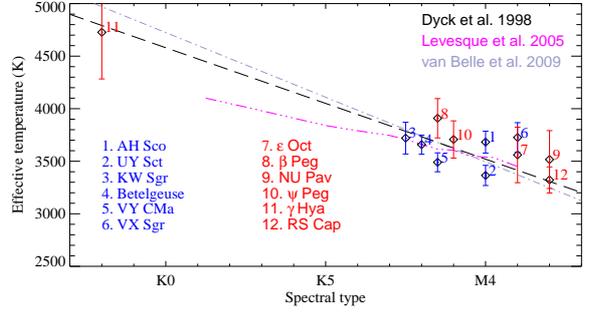}
\caption{Effective temperature versus spectral type of our sources in this paper, RS~Cap from Mart{\'{\i}}-Vidal et al. \cite{Marti2011}), and RSGs in Arroyo-Torres et al. (\cite{Arroyo2013}), VY~CMa from Wittkowski et al. (\cite{Witt2012}), Betelgeuse from Ohnaka et al. (\cite{Ohnaka2011}), and VX~Sgr from Chen et al. (\cite{Chen2007}) and Chiavassa et al.
(\cite{Chiavassa2010}). We also show the scales of  Dyck et al. (\cite{Dyck1998}), Levesque et al. (\cite{Levesque2005}), and van Belle et al. (\cite{Belle2009}).}
\label{Teff_sp_R}
\end{figure}

Figure \ref{Teff_sp_R} shows effective temperature vs. spectral type for our targets. We also include in the figures the AGB star RS~Cap from Mart{\'{\i}}-Vidal et al. (\cite{Marti2011}), and the following sets of RSGs: AH~Sco, UY~Sct, and Kw~Sgr from Arroyo-Torres et al. (\cite{Arroyo2013}), VY~CMa from Wittkowski et al. (\cite{Witt2012}), Betelgeuse from Ohnaka et al. (\cite{Ohnaka2011}), and VX~Sgr from Chen et al. (\cite{Chen2007}) and Chiavassa et al. (\cite{Chiavassa2010}). To calculate the effective temperature of RS~Cap we used the angular diameter from Mart{\'{\i}}-Vidal et al. (\cite{Marti2011}) and the bolometric flux from Dyck et al. (\cite{Dyck1998}).

For comparison, Fig \ref{Teff_sp_R} includes the calibrations of the effective temperature scale by Dyck et al. (\cite{Dyck1998}) for cool giants stars and van Belle et al. (\cite{Belle2009}) for cool giants stars and RSG stars. We also show the effective temperature scale by Levesque et al. (\cite{Levesque2005}) only for RSGs. Given our observational uncertainties, the cool giant stars agree with the calibration of these effective temperature scales.

\subsection{HR-diagram}

\begin{figure}
\centering
\includegraphics[width=0.9\hsize]{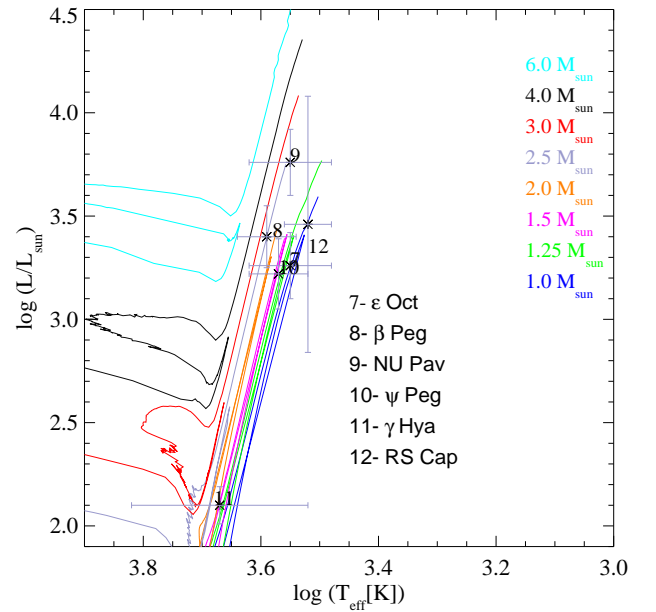}
\caption{Location of $\epsilon$~Oct, $\beta$~Peg, NU~Pav, $\psi$~Peg, and $\gamma$~Hya in the HR diagram using our  determination of the fundamental parameters. We also show the position of RS~Cap by Mart{\'{\i}}-Vidal et al. (\cite{Marti2011}). The positions of the stars are compared with evolutionary tracks from Lagarde et al. (\cite{Lagarde2012}) for masses of 1.0\,$M_{\odot}$, 1.25\,$M_{\odot}$, 1.5\,$M_{\odot}$, 2.0\,$M_{\odot}$, 2.5\,$M_{\odot}$, 3.0\,$M_{\odot}$, 4.0\,$M_{\odot}$, and 6.0\,$M_{\odot}$.}
\label{diagram_HR_Lagarde}
\end{figure}

Figure \ref{diagram_HR_Lagarde} shows the positions of $\epsilon$~Oct, $\beta$~Peg, NU~Pav, $\psi$~Peg, and $\gamma$~Hya in the Hertzsprung-Russell (HR) diagram, together with the evolutionary tracks from the model STAREVOL (Lagarde et al. \cite{Lagarde2012}). We used the model for which the transport processes in radiative zones are performed by thermohaline mixing and rotation-induced mixing. Thermohaline mixing occurs when material of high mean molecular weight lies on top of material of low mean molecular weight, a situation that is unstable against a blob of material moving downward, causing the mixture. This process is developed along the red giant branch in low-mass stars and on the early-AGB in intermediate-mass stars (more information in Lagarde et al. \cite{Lagarde2012}). We also show the position of RS~Cap from Mart{\'{\i}}-Vidal et al. (\cite{Marti2011}). We represent evolutionary tracks with masses between 1.0\,$M_{\odot}$ and 6\,$M_{\odot}$ and solar metallicity. All targets are close to the red limit of these tracks (Hayashi limit). $\epsilon$~Oct, $\psi$~Peg, $\gamma$~Hya, and RS~Cap are close to the evolutionary tracks with initial masses between 1\,$M_{\odot}$ and 3\,$M_{\odot}$ (the age of $\gamma$~Hya is younger than the others sources). $\beta$~Peg is consistent with tracks of masses between 1.25\,$M_{\odot}$ and 4\,$M_{\odot}$ and NU~Pav with tracks of masses between 2.5\,$M_{\odot}$ and 6\,$M_{\odot}$.

Additionally, we compare the position of our stars with the evolutionary tracks from Ekstr\"om et al. (\cite{Ekstrom2012}). We used the evolutionary tracks in Arroyo-Torres et al. (\cite{Arroyo2013}) for comparison with a sample of RSGs (AH~Sco, UY~Sct, and KW~Sgr) because they were a good choice for RSGs. In this work, we show both samples for comparison purposes. We also show the position of RS~Cap from Mart{\'{\i}}-Vidal et al. (\cite{Marti2011}), VY~CMa from Wittkowski et al. (\cite{Witt2012}), Betelgeuse from Ohnaka et al. (\cite{Ohnaka2011}), and VX~Sgr from Chen et al. (\cite{Chen2007}) and Chiavassa et al.(\cite{Chiavassa2010}). With these tracks, we observe that our stars are located considerably to the right of the models, unlike with the Lagarde evolutionary tracks. The positions of $\epsilon$~Oct, $\beta$~Peg, and $\psi$~Peg are close to the evolutionary tracks with initial masses of 5 or 7\,$M_{\odot}$. NU~Pav is consistent with tracks of masses of 7 or 9\,$M_{\odot}$, $\gamma$~Hya with mass of 3\,$M_{\odot}$, and RS~Cap with masses of 5-9\,$M_{\odot}$. In all cases the tracks are shown without/with rotation. The position of the RSGs are close to tracks with masses between 20\,$M_{\odot}$ and 40\,$M_{\odot}$. As expected, RSGs are much more massive and luminous than cool giants stars for the same effective temperatures. 

The STAREVOL model is complementary to the Ekstr\"om model for low-mass stars. The Ekstr\"om model is computed only until the helium flash at the RGB tip and does not include thermohaline mixing. Both models use almost the same assumptions and input physics: convection, opacity, mass loss, and nuclear reaction rates. The initial abundances are also similar, although the STAREVOL model considers more species (for more information see Lagarde et al. \cite{Lagarde2012}).

\begin{figure}
\centering
\includegraphics[width=0.9\hsize]{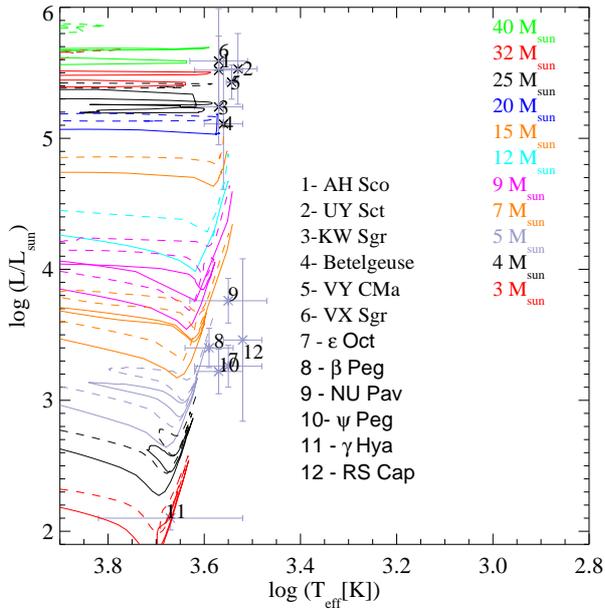}
\caption{Location of $\epsilon$~Oct, $\beta$~Peg, NU~Pav, $\psi$~Peg, and $\gamma$~Hya in the HR diagram using our determination of the fundamental parameters. We also show the position of RS~Cap by Mart{\'{\i}}-Vidal et al. (\cite{Marti2011}), and the RSGs studied in Arroyo-Torres et al. (\cite{Arroyo2013}), VY~CMa from Wittkowski et al. (\cite{Witt2012}), Betelgeuse from Ohnaka et al. (\cite{Ohnaka2011}), and VX~Sgr from Chen et al. (\cite{Chen2007}) and Chiavassa et al.(\cite{Chiavassa2010}). The positions of the stars are compared with evolutionary tracks from Ekstr\"om et al. (\cite{Ekstrom2012}) for masses of 3\,$M_{\odot}$, 4\,$M_{\odot}$, 5\,$M_{\odot}$, 7\,$M_{\odot}$, 9\,$M_{\odot}$, 12\,$M_{\odot}$, 15\,$M_{\odot}$, 20\,$M_{\odot}$, 25\,$M_{\odot}$, 32\,$M_{\odot}$, and 40\,$M_{\odot}$. The solid lines are models without rotation, the dashed lines with rotation.}
\label{diagram_HR}
\end{figure}


\section{Conclusions}

Our spectro-interferometric near-infrared observations of $\epsilon$~Oct, NU~Pav, $\psi$~Peg, and $\gamma$~Hya show that synthetic visibilities from hydrostatic atmospheric models are consistent with the observations, concluding that their atmospheres can be modeled with a limb-darkened disk. In $\epsilon$~Oct,  NU~Pav, and $\gamma$~Hya, the uniform disk diameter is constant across the band, and the CO bandheads present a similar size to that of the continuum. On the other hand, the data of $\psi$~Peg show a low increase in the CO band, similar to the one obtained in the model. According to these results, the atmospheres of $\epsilon$~Oct, NU~Pav, $\psi$~Peg, and $\gamma$~Hya are compatible with hydrostatic atmospheres and the role of pulsation does not seem to be important. However, the data from $\beta$~Peg (at least in the 2012 June epoch) show a layer (possibly of H$_{2}$O) that is not modeled by PHOENIX, but CO bands similar to those modeled with PHOENIX. The uniform disk diameter of the star at the CO band increases about 5.3\% with respect to the continuum (less than the 14\% increase of diameter observed in RS~Cap).

We used the continuum near 2.2\,$\mu$m, which is free from molecular band contamination, to estimate the angular diameter of the targets (see Table \ref{angular_diam}). We also estimated fundamental parameters such as the luminosity, Rosseland radius, and temperature (shown in Table \ref{fund_parameters}).

Finally, we located each of our targets in the HR diagram using the effective temperature and the luminosity calculated from the Rosseland angular diameter, the bolometric flux, and the distance. In the HR diagram, we also showed the evolutionary tracks from Lagarde et al. (\cite{Lagarde2012}). The positions of the stars in this HR diagram are close to the Hayashi limit. Their positions are close to evolutionary tracks corresponding to stars of initial masses between 1.0\,$M_{\odot}$ and 3\,$M_{\odot}$ ($\epsilon$~Oct, $\psi$~Peg, $\gamma$~Hya, and RS~Cap), between 1.25\,$M_{\odot}$ and 4\,$M_{\odot}$ ($\beta$~Peg), and between 2.5\,$M_{\odot}$ and 6\,$M_{\odot}$ ($\gamma$~Hya). We also compared the position of our stars with the evolutionary tracks from Ekstr\"om et al. (\cite{Ekstrom2012}). The STAREVOL model fits the positions of our stars in the HR diagram better than the Ekstr\"om model. This is probably because the STAREVOL model is designed for low-mass stars on the red giant branch and for intermediate-mass stars on the early-AGB. It is complementary to the Ekstr\"om model for low- and intermediate- mass stars.

\begin{acknowledgements}
This research has made use of the AMBER data reduction package of the Jean-Marie Mariotti Center and the SIMBAD database operated at CDS, Strasbourg, France. BAT, JCG and JMM acknowledge support by the Spanish Ministry of Science and Innovation though the grants AYA2009-13036-C02-02 and AYA2012-38491-C02-01.

\end{acknowledgements}

\object{epsilon Oct} 
\object{beta Peg} 
\object{NU Pav}
\object{psi Peg}
\object{gamma Hya}

\listofobjects

\end{document}